\documentclass[twocolumn,showpacs,preprintnumbers,amsmath,amssymb,nofootinbib]{revtex4}
\usepackage{graphicx}% Include figure files
\usepackage{dcolumn}% Align table columns on decimal point
\usepackage{bm}% bold math

\usepackage{amsmath}    % need for subequations
\usepackage{graphicx}   % for figures

\begin{document}

\title{Dynamics of nuclear single-particle structure in covariant
theory of particle-vibration coupling: from light to
superheavy nuclei}
%Particle-vibration coupling in covariant density functional
%theory: the impact on single particle motion in the nuclei from
%$^{56}$Ni up to superheavy ones.}

%Lines break automatically or can be forced with \\
\author{E. Litvinova}
\affiliation{GSI Helmholtzzentrum f\"{u}r Schwerionenforschung,
64291 Darmstadt, Germany}
\affiliation{Institut f\"{u}r Theoretische Physik,
Goethe-Universit\"{a}t, 60438 Frankfurt am Main, Germany}
\author{A.V. Afanasjev}
\affiliation{Department of Physics and Astronomy, Mississippi State
University, MS 39762}
\date{\today}

\begin{abstract}
The impact of particle-vibration coupling and polarization effects 
due to deformation and time-odd mean fields on single-particle 
spectra is studied systematically in doubly magic nuclei from 
low mass $^{56}$Ni up to superheavy ones. Particle-vibration
coupling is treated fully self-consistently within the framework
of relativistic particle-vibration coupling model. Polarization
effects due to deformation and time-odd mean field induced by
odd particle are computed within covariant density functional 
theory. It has been found that among these contributions the 
coupling to vibrations makes a major impact on the single-particle 
structure. The impact of particle-vibration coupling and polarization
effects on calculated single-particle spectra, the size of the 
shell gaps, the spin-orbit splittings and the energy splittings in 
pseudospin doublets is discussed in detail; these physical observables 
are compared with experiment. Particle-vibration coupling has to be 
taken into account when model calculations are compared with experiment 
since this coupling is responsible for observed fragmentation of 
experimental levels; experimental spectroscopic factors are reasonably 
well described in model calculations.
\end{abstract}

\maketitle

%%%%%%%%%%%%%%%%%%%%%%%%%%%%%%%%%%%%%%%%%%%%%%%%
\section{Introduction}
%%%%%%%%%%%%%%%%%%%%%%%%%%%%%%%%%%%%%%%%%%%%%%%%

   The covariant density functional theory (CDFT) \cite{Rei.89,Rin.96,VRAL.05}
is one of standard tools of nuclear theory which offers considerable potential for
further development. Built on Lorentz covariance and the Dirac equation, it provides
a natural incorporation of spin degrees of freedom~\cite{Rei.89,Rin.96} and an
accurate description of spin-orbit splittings~\cite{Rin.96} (see also Fig. 2 in
Ref.\ \cite{BRRMG.99}), which has an essential influence on the underlying shell
structure. Note that the spin-orbit interaction is a relativistic effect, which
arises naturally in the CDFT theory. Lorentz covariance of the CDFT equations
leads  to the fact that time-odd mean fields of this theory are determined as
spatial components of Lorentz vectors and therefore coupled with the same constants
as time-like components~\cite{AA.10} which are fitted to ground state properties
of finite nuclei. In addition, pseudo-spin symmetry finds a natural explanation in the
relativistic framework~\cite{Gin.97}. CDFT in its different incarnations both
on the mean field and beyond mean field levels provides succesful description
of many properties of ground state and excited configurations in nuclei
\cite{VRAL.05}.

  However, the majority of applications of CDFT have been focused on
collective properties of nuclei. There are only few features on nuclear
systems, strongly dependent on single-particle degrees of freedom, which
have been addressed in the CDFT studies on the mean field level and compared
with experiment. These are the single-particle properties in spherical and
deformed nuclei, spin-orbit splittings in spherical nuclei, magnetic moments
in the $A\pm 1$ neighbours of the $^{16}$O nucleus ($A=16$) \cite{HR.88},
and alignment  properties of single-particle orbitals in rotating nuclei
\cite{ALR.98}\footnote{Effective alignments of compared rotational bands,
which sensitively depend on both the alignment properties of single-particle
orbital by which two bands differ and polarization effects induced by the
particle in this orbital, are in average better reproduced in the CDFT
calculations than in the cranked Nilsson-Strutinsky calculations based on
phenomenological Nilsson potential, see comparisons presented in Refs.\
\cite{ARR.99,AR.00,AF.05}. This is despite the fact that no single-particle
information has been used in the fit of the CDFT parameters, while the
parameters of the Nilsson potential are fitted to experimental single-particle
energies (see Ref.\ \cite{Beng85}).}.

%%%%%%%%%%%%%%%%%%%%%%%%%%%%%%%%%%%%%%%%%%%%%%%%%%%%%%%%%%%%%%%%%%%%%%%%%%%%%%
\begin{figure*}[ht]
\includegraphics[width=16.0cm]{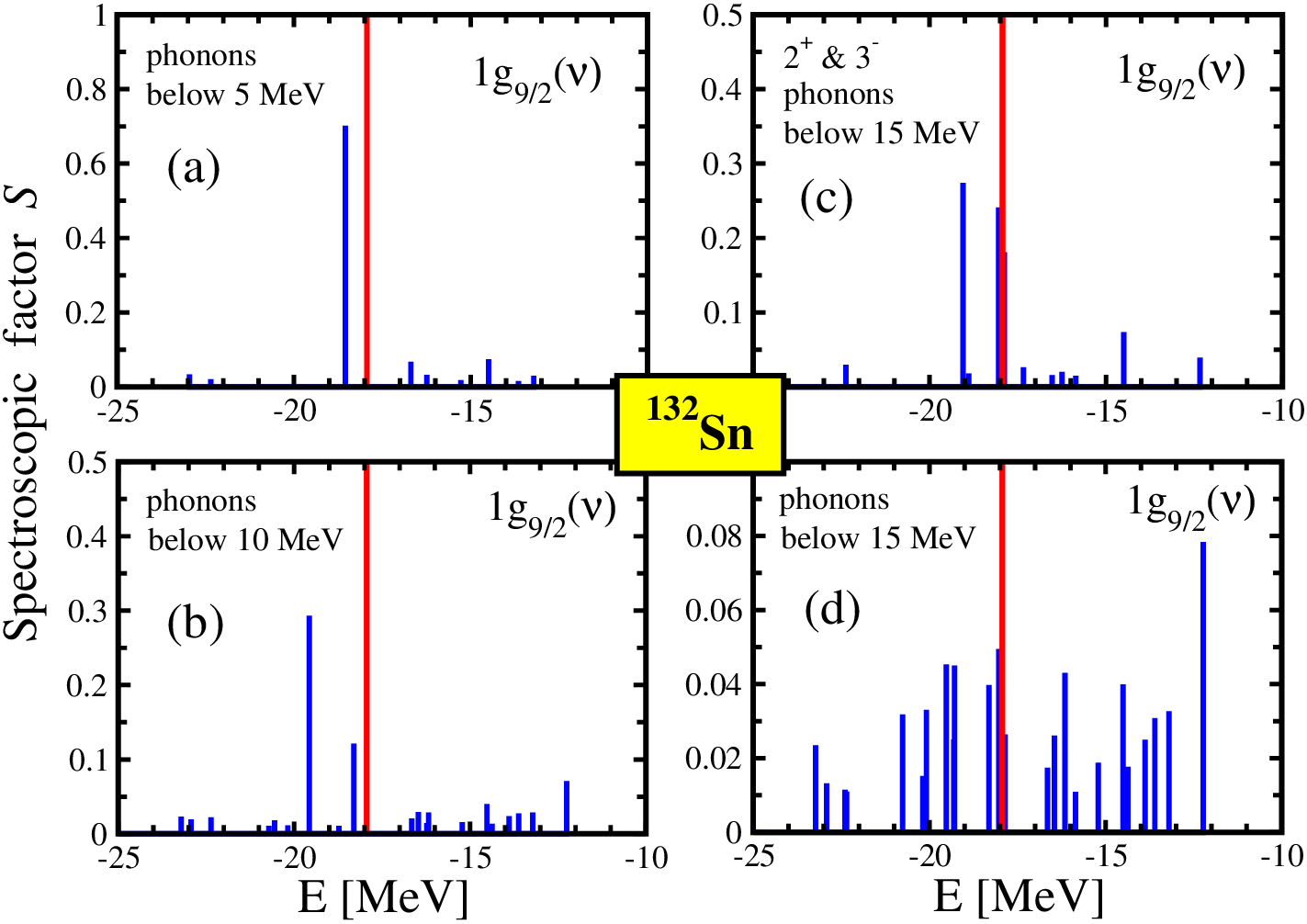}
%\vspace{0.8cm}
\caption{ (Color online) Single-particle strength distributions of
the neutron 1g$_{9/2}$ state in $^{132}$Sn calculated within the PVC
model with different numbers of the phonon modes. Red bars show the
initial mean field 1g$_{9/2}$ neutron state.}
\label{n1g92}
\end{figure*}
%%%%%%%%%%%%%%%%%%%%%%%%%%%%%%%%%%%%%%%%%%%%%%%%%%%%%%%%%%%%%%%%%%%%%%%%%%%%%%

 The calculated single-particle spectra of spherical nuclei are frequently
compared with experiment (see, for example, Refs.\ \cite{Rei.89,Rin.96}).
However, only in Refs.\ \cite{RBRMG.98,BRRMG.99} deformation polarization
effects and time-odd mean fields are included in the calculations which makes
them more realistic.
 The spin-orbit splittings, as extracted from the single-particle energies
obtained in spherical CDFT calculations, reproduce the experimental spin-orbit
splittings fairly well, although there are deviations up to 20\% of absolute
value of the splitting (see Figs. 11-12 and Table 4 in Ref.\ \cite{RBRMG.98}
and Fig. 2 and Sect. IVB in Ref.\ \cite{BRRMG.99}) which only weakly depend on
the RMF parametrization. Note that the comparison of Ref.\ \cite{BRRMG.99}
includes the uncertainty of the spin-orbit splittings due to deformation
polarization effects and time-odd mean fields as they are found in Ref.\
\cite{RBRMG.98}.  Ref.\ \cite{BRRMG.99} clearly shows that spin-orbit splittings
are better described in CDFT as compared with Skyrme energy density functional
theory (SEDFT) despite the fact that no single-particle information (contrary
to SEDFT) has been used in the fit of the RMF Lagrangian.

      These comparisons, however, do not reveal the accuracy of the description
of the single-particle states because the particle-vibration coupling (PVC), which
can affect considerably the energies of single-particle states in odd-mass
nuclei \cite{QF.78,RS.80,MBBD.85,LR.06,BBBBCV.06}, has been neglected. 
The modification
of the quasiparticle states by particle-vibration coupling is weaker in deformed
nuclei \cite{Sol-book,Sol-book2} since the surface vibrations are more fragmented
(less collective) than in spherical nuclei \cite{Sol-book,BM}. As a consequence,
the corrections to the energies of quasiparticle states in odd nuclei due to
particle-vibration coupling are expected to be less state-dependent in deformed
nuclei. Hence the comparison between experimental and calculated energies of
single-particle
states is expected to be less ambiguous in deformed nuclei as compared with
spherical ones \cite{MBBD.85,BM}, at least at low excitation energies, where
vibrational admixtures in the wave functions are small. The analysis of the
energies of one-quasiparticle deformed states in actinide region, performed
within the relativistic Hartree-Bogoliubov (RHB) approach in Refs.\
\cite{AKFLA.03,SA.11}, reveals that while the majority of the states are described
with an accuracy better than 0.5 MeV, there are a number of states which deviate
from experiment by as much as 1 MeV.

%%%%%%%%%%%%%%%%%%%%%%%%%%%%%%%%%%%%%%%%%%%%%%%%%%%%%%%%%%%%%%%%%%%%%%%%%%%%%%
\begin{figure*}[ht]
\includegraphics[width=16.0cm]{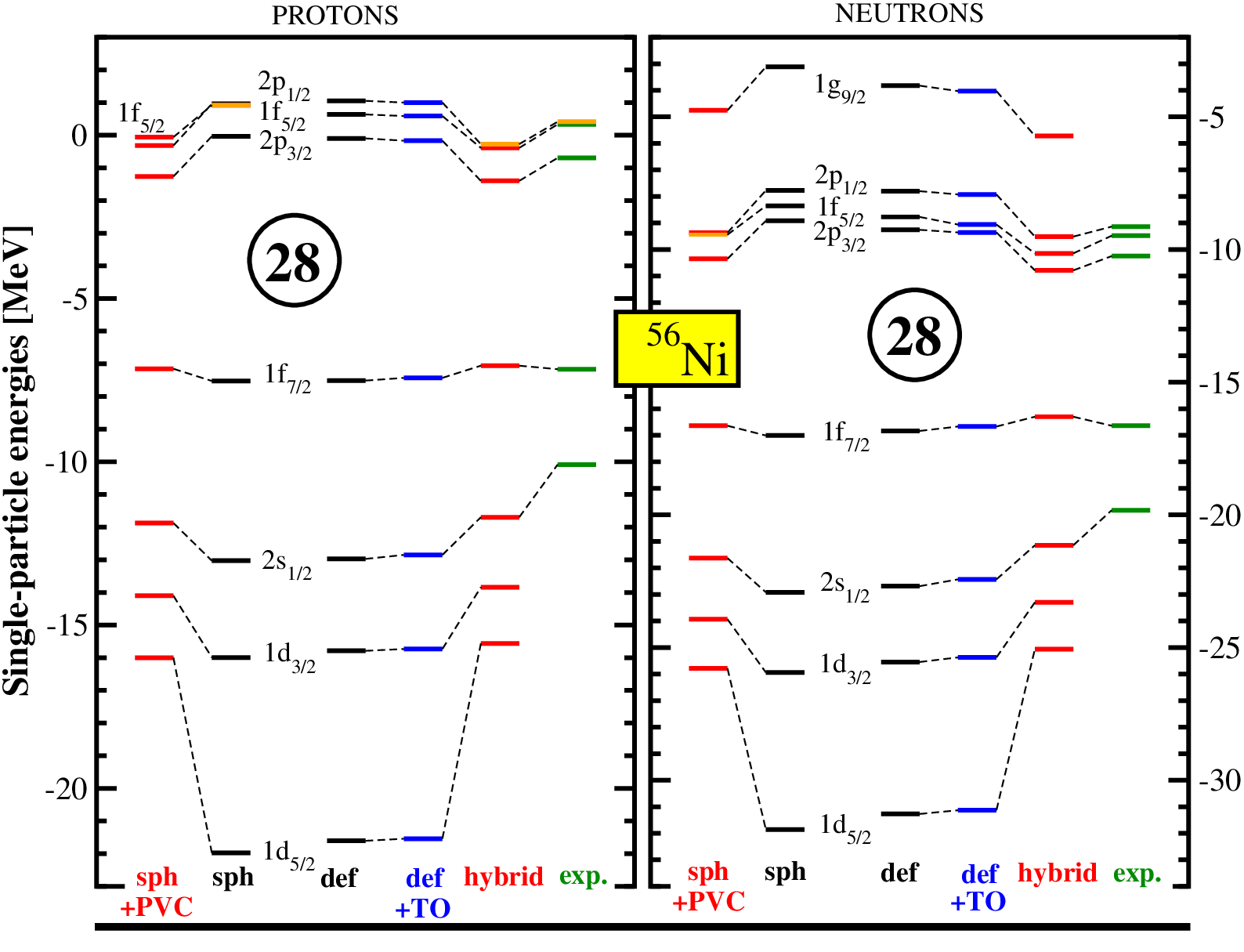}
\caption{(Color online) Spectra of $^{56}$Ni and its
neighboring odd nuclei. Column 'sph' shows the single-particle
spectra obtained in spherical RMF calculations of $^{56}$Ni. 
Column 'sph+PVC' shows the spectra obtained in spherical
calculations within the PVC model. Columns ``def'',
``def+TO'', ``hybrid'' and ``exp'' show one-nucleon
separation energies defined according to Eqs.\ (\ref{Eq-part}) and
(\ref{Eq-hole}). Column ``def'' is based on the results of triaxial
CRMF calculations with no TO mean fields. These fields are included
in the calculations the results of which are shown in column
``def+TO''. The corrections due to PVC are added in column
``hybrid''. The experimental single-particle energies are 
displayed in column ``exp''; they are based on the data of Refs.\ 
\cite{AWT.03} (masses of ground states), \cite{Ni55} ($^{55}$Ni), 
\cite{Co55} ($^{55}$Co), \cite{Ni57,TKS.96} ($^{57}$Ni) and 
\cite{TKS.96} ($^{57}$Cu). In order to distinguish overlapping 
levels, orange and then maroon colors are used for the levels in 
addition to their standard color used in a given column. See text for 
more details.}
\label{Ni56-fig}
\end{figure*}
%%%%%%%%%%%%%%%%%%%%%%%%%%%%%%%%%%%%%%%%%%%%%%%%%%%%%%%%%%%%%%%%%%%%%%%%%%%%%%

  The question arises of which features of the DFT are most
decisive for the single-particle structure. The most crucial
ingredients in this respect are
\begin{itemize}
\item
{\it the effective nucleon mass and its radial dependence which
determines the level density near the Fermi surface}. It is well
known that the one-(quasi-) particle spectra calculated on the mean
field level are less dense than in experiment both at spherical
\cite{QF.78,RS.80,MBBD.85,LR.06} and deformed \cite{AKFLA.03,SA.11}
shapes.  The average level density of the single-particle states on
the mean field level is related to the effective mass (Lorentz mass
in the notation of Ref.\ \cite{JM.89} for the case of CDFT theory)
of the nucleons at the Fermi surface $m^*_L(k_F)/m$. The CDFT is
characterized by a low effective mass $m^*_L(k_F)/m \approx 0.66$
\cite{BRRMG.99}. As a consequence, this low effective mass of CDFT
has a bigger impact on the single-particle spectra than the effective
masses of non-relativistic DFT which are typically larger. Note that
at spherical shape the inclusion of particle-vibrational coupling
brings the calculated level density closer to experimental one
characterized by an effective mass $m^*(k_F)/m\approx 1.0$
\cite{LR.06}.

\item
  {\it the spin-orbit potential which determines the energetic distance
of the spin-orbit partners}. It is well known that CDFT describes
rather well spin-orbit splittings \cite{BRRMG.99}. However, these results
have been obtained on the mean field level, and it is necessary to test
how they will be affected by the inclusion of particle-vibrational
coupling. In CDFT theory the Dirac effective mass $m^*_D(k_F)/m$ is closely
related to the effective spin-orbit single-nucleon potential, and empirical
energy spacings between spin-orbit partner states in finite nuclei determine
a relatively narrow interval of allowed values: $0.57\leq m^*_D(k_F)/m \leq 0.61$
\cite{NVR.08}. This requirement restricts considerably the possible
parametrizations of the RMF Lagrangian on the mean field level, and thus
it would be interesting to see whether the inclusion of PVC will loosen
this requirement and thus allow wider range of effective masses in the
CDFT.

\item
{\it the density dependence of the effective potential and effective
mass which has an influence on the relative position of the states
with different orbital angular momentum $l$.} The investigation
of Ref.\ \cite{RBRMG.98}, covering spherical odd nuclei bordering
doubly magic nuclei clearly showed that the relative placement of
the states with different orbital angular momenta $\it l$ is not so
well reproduced on the mean field level in CDFT. The problems occur
predominantly in connection with the states of high $\it l$ which hints
that the surface profile of the mean-field and kinetic terms are
involved. Microscopic consideration in many-body theory indicates 
that the effective mass has a pronounced surface profile which is 
insufficiently parametrized in the present mean-field models 
\cite{MBBD.85}. Thus, it is important to see whether the inclusion 
of PVC will improve the description of the relative positions of 
the states with different orbital angular momentum $l$.

\end{itemize}

  In Ref. \cite{LR.06}, a relativistic particle-vibration coupling 
model has been developed and the calculations based on the NL3 
\cite{NL3} parametrization of the CDFT have been performed for the 
doubly magic $^{208}$Pb nucleus. In this work we present a more 
systematic investigation within this model including the calculations 
covering the nuclei from light $^{56}$Ni and medium-mass  $^{100}$Sn, 
$^{132}$Sn to heavy  $^{208}$Pb and superheavy $^{292}$120
doubly magic nuclei using improved (as compared with NL3) NL3* 
\cite{NL3*} parametrization of the CDFT. This investigation covers 
neutron-deficient near proton drip-line nuclei ($^{56}$Ni and  
$^{100}$Sn), neutron-rich $^{132}$Sn nucleus, and 
the $^{208}$Pb  nucleus located at the beta-stability line.
In addition, the polarization effects in odd-mass nuclei due to 
deformation and time-odd mean fields are treated by means of cranked 
relativistic mean field (CRMF) calculations. 
 
   The article is organized as follows. The method of comparison 
of experimental and theoretical single-particle levels is discussed 
in Sec.\ \ref{Exp-vs-th}. Section \ref{Sec-theory} presents 
relativistic particle-vibration model and cranked relativistic mean 
field theory and their details related to the study of single-particle 
states. The comparison of model calculations with experiment is 
presented in Sec.\ \ref{Res-disc}. The accuracy of the description 
of single-particle spectra and spectroscopic factors, the impact of 
particle-vibration coupling on shell gaps, spin-orbit splittings and 
pseudo-spin doublets as well as the role of polarization effects in 
odd-mass nuclei due to deformation and time-odd mean fields are 
discussed in detail. Finally, Sec.\ \ref{Sec-final} reports the main 
conclusions of our work.

%%%%%%%%%%%%%%%%%%%%%%%%%%%%%%%%%%%%%%%%%%%%%%%%%%%%%%%%%%%%%%%%%%%
\section{Experimental versus theoretical single-particle energies}
\label{Exp-vs-th}
%%%%%%%%%%%%%%%%%%%%%%%%%%%%%%%%%%%%%%%%%%%%%%%%%%%%%%%%%%%%%%%%%%%

  The experimental information about single-particle level structure
of even-even doubly magic nucleus with the proton number $Z_0$ and neutron
number $N_0$ originates from the odd-mass $(Z_0\pm 1)$ and $(N_0 \pm 1)$ 
nuclear neighbors. In odd-mass nuclei, the polarization effects due to 
deformation and time-odd (TO) mean fields\footnote{The name {\bf 
nuclear magnetism} is frequently used in the CDFT to describe the 
effect of time-odd mean fields induced by magnetic potential of the
Dirac equation since this potential has the structure of a magnetic
field \cite{KR.89,AA.10}.} induced by odd particle play an 
important role (see Sec.\ \ref{Pol-eff}). In addition, the particle-vibration 
coupling modifies the energies and the structure of the levels (Secs.\ 
\ref{SP-acc} and \ref{Spec-fac}); neither of them remains purely 
single-particle in nature. In order to describe all these effects we use 
a hybrid model (labelled as ``hybrid'' below), in which the polarization
effects due to deformation and TO mean fields are treated on
the mean field level within cranked relativistic mean field (CRMF)
approach (see Sect.\ \ref{Mean-field}), while the corrections due to
the PVC are treated within relativistic particle-vibration coupling 
model (see Sect.\ \ref{Beyond-mean-field}). The need for a such hybrid 
model is dictated by the fact that the PVC model is not variational in nature 
and thus it does not allow to calculate polarization effects due to odd 
particle. Since the polarization effects and the corrections due to the PVC have 
a different origin, they are treated as additive quantities in the hybrid model
which is a reasonable approximation. The details of the hybrid model are 
discussed in Sec.\ \ref{Sec-theory}.

 The experimental and calculated energies of the particle [$\varepsilon({\rm particle})$]
and hole [$\varepsilon({\rm hole})$] states closest to the Fermi level are determined from 
the difference of the binding energies of the core [$B({\rm core})$]
and the corresponding adjacent odd nuclei [$B({\rm core+nucleon})$
and $B({\rm core-nucleon})$] as \cite{RBRMG.98,IEMSF.02} 
\begin{equation}
\varepsilon({\rm particle})=B({\rm core})-B({\rm core+nucleon})
\label{Eq-part}
\end{equation}
and
\begin{equation}
\varepsilon({\rm hole})=B({\rm core-nucleon})-B({\rm core}).
\label{Eq-hole}
\end{equation}
 These quantities correspond to one-particle removal energies. We will 
call them as single-particle energies for simplicity of discussion. Doubly 
magic spherical nuclei $^{56}$Ni, $^{100,132}$Sn, and $^{208}$Pb are used 
as the cores in our analysis. Their ground states are not affected by PVC.  The 
experimental energies of particles and holes defined according to these 
equations include the polarization effects due to deformation and TO mean 
fields induced by odd particle or hole as well as the energy corrections 
due to particle-vibration coupling in odd-mass nuclei. Thus, their comparison 
with the results of ``hybrid'' calculations is straightforward.

   On the other hand, there are some unresolved questions when these
energies are compared with pure mean field calculations as it is
done frequently in the literature. The problem is related to the
fact that experimental single-particle levels and their energies and 
structure are affected by the particle-vibration coupling which is 
neglected on the mean field level. Each mean field state $k$ with
the energy $\varepsilon_k$ is split into many levels due to particle-vibration 
coupling, so the single-particle strength is fragmented over many levels. 
In the diagonal approximation for the nucleonic self-energy, these levels 
have the same quantum numbers as the original mean field state $k$, but 
different energies $\varepsilon^{\nu}_{k}$ and spectroscopic factors 
$S^{\nu}_{k}$. Unlike the Hartree or Hartree-Fock approximations without 
pairing where the 
single-particle states are either fully occupied or empty, in the PVC 
model their occupation probabilities are the real numbers between zero 
and one, so that sum rule 
\begin{equation}
\sum\limits_{\nu}S^{\nu}_{k} = 1
\label{sum_rule}
\end{equation}
is satisfied. For the states in the vicinity of the Fermi surface one usually 
obtains  one dominant level with $0.5 \leq S^{\nu}_{k} \leq 1.0$ and many other 
levels with small $S^{\nu}_{k}$. For the states far from the Fermi surface 
one observes very strong splitting over many levels with much smaller and 
comparable spectroscopic factors, see the detailed analysis in Ref.\ 
\cite{LR.06} for $^{208}$Pb.

  This discussion clearly indicates that some procedure has to be 
employed in order to extract pure (or ``bare'') single-particle 
energies from experimental data which have to be compared with 
mean field single-particle energies $\varepsilon_k$. Two such procedures 
have been discussed in the literature.

  In the first procedure, the energy of the dominant fragment is 
corrected by the energy shift due to particle-vibration coupling 
leading to a ``bare'' single-particle energy. Such energies 
of the doubly-magic nuclei are extracted in some publications, 
see, for example, Ref.\ \cite{TKS.96}. However, this procedure 
relies on the choice of particle-vibration coupling model.
However, different PVC models lead to different energy shifts 
(see Table 1 in Ref.\ \cite{BCS.10}). In addition, such procedure 
frequently neglects polarization effects due to deformation and 
TO mean fields which are again model dependent. For example, there 
is no time-odd mean fields in the phenomenological models based on 
the Nilsson or Woods-Saxon potentials.

 In the second procedure, the center of gravity of the 
fragmented levels with a given quantum number $j^{\pi}$ 
($j$ is the total angular momentum and $\pi$ is a parity)
and energies $\varepsilon^{\nu}_{k}$ is obtained via 
weighted-average \cite{B.70,DSPMRF.10} procedure
\begin{eqnarray}
\varepsilon_k^{grav} = 
 \left[ \sum_\nu S_k^{\nu} \cdot \varepsilon^{\nu}_k \right]
 / 
\left[ \sum_\nu S_k^{\nu} \right]
\end{eqnarray}
 This energy is then associated with a ``bare'' single-particle
energy. This expression is fulfilled exactly in the PVC 
model discussed in Sec.\ \ref{Beyond-mean-field} and it is 
satisfied with high accuracy in this model numerical 
implementation. The normalization in this procedure (which is 
omitted in some publications) is used to minimize the uncertainties 
due to large experimental errors on spectroscopic factors, since 
these errors lead to the fact that the sum rule $\sum_{\nu} S_k^{\nu}=1$ 
is poorly fulfilled in the majority of 
experiments. The application of this procedure requires that 
all $\nu$-levels, to which the single-particle state $k$ is fragmented, 
are identified in the experiment and that spectroscopic factors are
measured accurately. These conditions are definetely not satisfied
in the $^{56}$Ni and $^{100,132}$Sn nuclei.

   The situation is somewhat better in $^{208}$Pb, for which the 
$\varepsilon_k^{grav}$ values have been extracted in Ref.\ \cite{DSPMRF.10} 
for a number of levels. However, even in this case the accuracy of the 
definition of $\varepsilon_k^{grav}$ is not known because of the number 
of the reasons listed below. The absolute values of spectroscopic 
factors are characterized by large ambiguities and depend strongly on the 
reaction employed in  experiment and the reaction model used in the 
analysis \cite{LTLHS.09,DSPMRF.10}. The spectroscopic factors in odd 
mass nuclei have been obtained in different reactions: for example, in the
(d,$^{3}$He) reaction for $^{207}$Tl \cite{207Tl},  ($\alpha$, t) for 
$^{209}$Bi \cite{209Bi}, ($^{3}$He, $\alpha$) \cite{207Pb-1} and $(d,t)$ 
\cite{207Pb-2} for $^{207}$Pb. Even in the same nucleus different reactions
give different spectroscopic factors (see, for example, Table I in Ref.\ 
\cite{207Pb-2} and Table \ref{Table-spec-1} in the current manuscript). In 
addition, the spins and parities of many high-lying fragments are uncertain 
(see, for example, Table 1 in Ref.\ \cite{207Tl}) and/or only the part of the sum 
rule strength has been observed in experiment (see, for example, Ref.\ \cite{209Bi}). 
As a consequence, it is not clear whether the complete set of the $\nu$-levels, 
to which the single-particle state $k$ is fragmented, and their spectroscopic
factors have been identified in experiment.
 
  In the light of this discussion it is clear that ``bare'' single-particle 
energies cannot be defined with controllable precision in the nuclei under
study. As a result, we compare observed experimental levels directly with
mean field calculations. Such an approach is frequently used in the 
literature (see, for example, Refs.\ \cite{Rin.96,RBRMG.98,IEMSF.02}). 
It  has clear limitations, but at least it allows  to see  (i)
whether the inclusion  of particle-vibration coupling improves the description 
of specific physical observables and (ii) whether the conclusions reached 
earlier on the mean field level are valid or not.

%%%%%%%%%%%%%%%%%%%%%%%%%%%%%%%%%%%%%%%%%%%%%%%%%%%%%%%%%%%%%%%%%%%%%%%%%%%%%%
\begin{figure*}[ht]
\includegraphics[width=16.0cm]{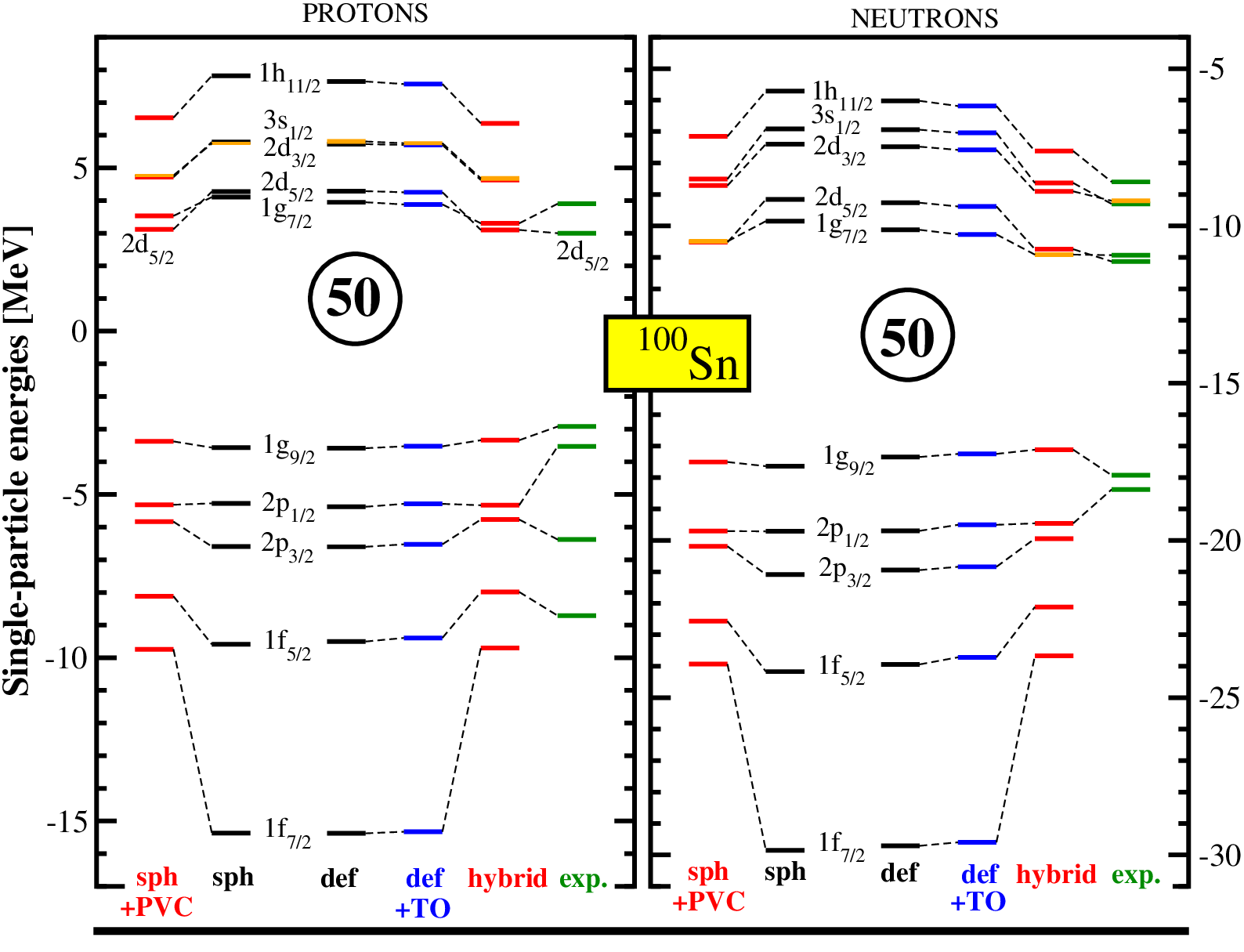}
%\vspace{1.0cm}
\caption{(Color online) The same as in Fig.\ \ref{Ni56-fig} but for the
spectra of $^{100}$Sn. Extrapolated ``experimental'' single-particle energies 
of proton and neutron spherical subshells are taken from Ref.\ \cite{Ext-1}.}
\label{Sn100-fig}
\end{figure*}
%%%%%%%%%%%%%%%%%%%%%%%%%%%%%%%%%%%%%%%%%%%%%%%%%%%%%%%%%%%%%%%%%%%%%%%%%%%%%%

%%%%%%%%%%%%%%%%%%%%%%%%%%%%%%%%%%%%%%%%%%%%%%%%%%%%%%%%%%
\section{Theoretical framework}
\label{Sec-theory}
%%%%%%%%%%%%%%%%%%%%%%%%%%%%%%%%%%%%%%%%%%%%%%%%%%%%%%%%%%

%%%%%%%%%%%%%%%%%%%%%%%%%%%%%%%%%%%%%%%%%%%%%%%%%%%%%%%%%%
\subsection{Mean field level: polarization effects}
\label{Mean-field}
%%%%%%%%%%%%%%%%%%%%%%%%%%%%%%%%%%%%%%%%%%%%%%%%%%%%%%%%%%

  Mean field results related to polarization effects due to deformation 
and TO mean fields in odd-mass nuclei have been obtained using 
cranked relativistic mean field (CRMF) theory \cite{KR.89,KR.93,AKR.96}.
Although this theory has initially been developed for the description 
of rotating nuclei, it is also able to describe the nuclear systems 
with broken  time-reversal symmetry in intrinsic frame at no rotation 
in which TO mean fields play an important role (see Ref.\ \cite{AA.10}).

  The application of the CRMF theory has the following advantages.
First, the CRMF computer code is formulated in the signature basis. 
As a result, the breaking of Kramer's degeneracy of the single-particle 
states in odd-mass nuclei is taken into account in a fully self-consistent 
way. This is important for an accurate description of TO mean fields (Ref.\ 
\cite{AA.10}). These fields have considerable impact on the moments 
of inertia in rotating nuclei \cite{AA.10b} and the fact that the
properties of such nuclei are well described in the CRMF code
(see Refs.\ \cite{VRAL.05,AA.10b} adds confidence to proper description
of TO mean fields in non-rotating systems \cite{AA.10}. Second, the CRMF 
computer code is formulated in three-dimensional cartesian coordinates
which allows to describe not only axially deformed but also triaxial
nuclear shapes. This is important for a proper description of polarization
effects due to deformation.

  It is well known fact that the description of pairing correlations in 
doubly magic nuclei and their close neighborhood is notoriously difficult 
and model dependent (Ref.\ \cite{AER.02}). The presence of large shell 
gaps leads to a pairing collapse in the models with the treatment of 
pairing on the BCS or HFB [Hartree-Fock-Bogoliubov] levels. This is seen, for 
example, in non-relativistic HFB calculations with the Gogny D1S force
of $^{100,132}$Sn in Ref.\ \cite{AER.02}. The calculations within 
the relativistic Hartree-Bogoliubov (RHB) framework of Ref.\ \cite{ARK.00} 
with the same force for pairing also show pairing collapse in the ground 
states of $^{56}$Ni, $^{100,132}$Sn and $^{208}$Pb. The pairing correlations 
are restored when the particle number projection is implemented. However, 
there is a substantial difference between different approaches (such as 
Lipkin-Nogami (LN) method of approximate particle number projection or 
projection after variation (PAV) approach) and variation after projection 
(VAP) approach; the latter represents the best approach for the treatment 
of pairing correlations \cite{RS.80}. The RHB+LN calculations with the Gogny
D1S force for pairing show very weak pairing in the nuclei of interest. 
The additional binding due to pairing (relatively to the results with no
pairing)  is only 0.333 MeV, 0.284 MeV, 0.493 MeV and 0.557 MeV in $^{56}$Ni, 
$^{100,132}$Sn and $^{208}$Pb, respectively. The blocking of single-particle 
orbital in odd-mass nucleus leads to the weakening of pairing correlations
and to frequent unphysical pairing collapse even in the RHB+LN calculations.
The physical quantities, such as given by Eqs.\ (\ref{Eq-part}) and (\ref{Eq-hole}), 
are relative in nature. Thus, we believe that in the case of weak pairing
they are sufficiently well described in unpaired calculations. As a 
consequence, we neglect pairing correlations in the CRMF calculations. Note
that they are also neglected in the PVC calculations.

  Similar to Ref.\ \cite{RBRMG.98}, binding energies of odd-$A$ nuclei 
were calculated by blocking the $(n_r+1,j^{\pi},j)$ state (amongst 
the $(n_r+1,j^{\pi},m)$ states) with the largest projection $m=j$ onto the 
symmetry axis. Here, the $n_r$ and $j$ are the single-nucleon radial and
total angular momentum quantum numbers. The spherical subshell label 
of this state  is used to label the configuration of odd mass nucleus since 
the deformations (induced by odd particle or hole) of immediate odd-$A$ 
neighbors of doubly magic nuclei are rather small.

%%%%%%%%%%%%%%%%%%%%%%%%%%%%%%%%%%%%%%%%%%%%%%%%%%%%%%%%%%%%%%%%%%%%%%%%%%%%%%
\begin{figure*}[ht]
\includegraphics[width=16.0cm]{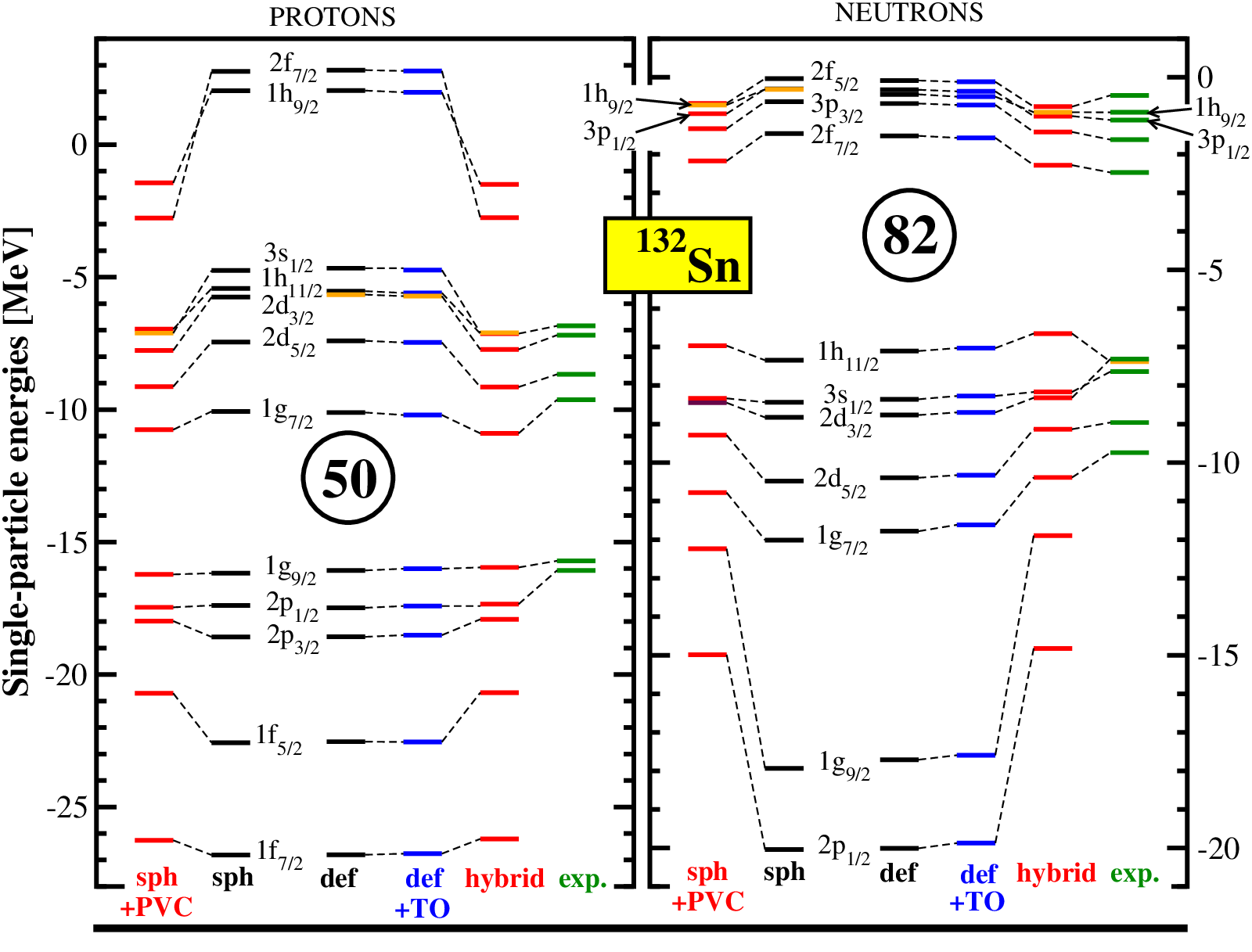}
%\includegraphics[width=16.0cm]{sn132-my-new-1.eps}
%\includegraphics[width=16.0cm]{sn132-my-new.eps}
%\vspace{1.0cm}
\caption{(Color online) The same as in Fig.\ \ref{Ni56-fig} but for the
spectra of $^{132}$Sn. The experimental single-particle energies are based
on the data of Refs.\ \cite{AWT.03} (masses of ground states), \cite{Sn131}
($^{131}$Sn), \cite{A131} ($^{131}$In), \cite{Sb133,Sb133-1} ($^{133}$Sb)
and \cite{Sn133} ($^{133}$Sn).}
\label{Sn132-fig}
\end{figure*}
%%%%%%%%%%%%%%%%%%%%%%%%%%%%%%%%%%%%%%%%%%%%%%%%%%%%%%%%%%%%%%%%%%%%%%%%%%%%%%

%%%%%%%%%%%%%%%%%%%%%%%%%%%%%%%%%%%%%%%%%%%%%%%%%%%%%%%%%%
\subsection{Beyond the covariant density functional theory:
single-particle spectra of nuclei}
\label{Beyond-mean-field}
%%%%%%%%%%%%%%%%%%%%%%%%%%%%%%%%%%%%%%%%%%%%%%%%%%%%%%%%%%

%%%%%%%%%%%%%%%%%%%%%%%%%%%%%%%%%%%%%%%%%%%%%%%%%%%%%%%%%%%%%%%%%%%%%%%%
\subsubsection{Particle-vibration coupling model of the time dependence
of the nucleonic self-energy}
%%%%%%%%%%%%%%%%%%%%%%%%%%%%%%%%%%%%%%%%%%%%%%%%%%%%%%%%%%%%%%%%%%%%%%%%

  The mean field approach, being a very useful and convenient
theoretical tool to describe finite nuclei, is, however, only a
static and local approximation to the many-body problem. The
non-locality (or, equivalently, momentum dependence) is implicitly
taken into account by an effective mass which is considerably
smaller than the bare nucleon mass. This enables one to reproduce
reasonably well the general properties of the single-particle motion 
in the nuclear potential well. However, the mean field approximations
based on the widely used Skyrme, Gogny as well as covariant density
functionals generate too stretched single-particle spectra around
the Fermi surface and often fail to describe the sequence of the
single-particle levels. Proper inclusion of the time dependence in
the nucleonic potential requires to go beyond the mean field
description which is static by construction.

 The underlying physics which is responsible for the time dependence
is related to many-body correlations, i.e. to the part of the
hamiltonian which cannot be expressed in terms of the one-body
density. In the medium mass and heavy nuclei the leading order
contribution of such correlations to the nucleonic self-energy comes
from coupling of the nucleonic motion to the surface and volume
vibrational modes (phonons). The most important collective
vibrational modes are generated by the nuclear oscillations of the
coherent nature. Taking into account the coupling of these
modes to the single-particle motion, the shell model acquires a
dynamic content. These surface and volume oscillations, especially
their low-lying modes, modify considerably the picture of the
single-particle motion in nuclei. The main assumption of the
quasiparticle-phonon coupling model is that two types of elementary
degrees of freedom -- one-quasiparticle (1-qp) and collective ones --
are coupled in such a way that configurations of 1-qp$\otimes$phonon
type with low-lying phonons strongly compete with simple 1-qp
configurations located close in energy or, in other words, that
quasiparticles can emit and absorb phonons with rather high
probabilities \cite{BM.75,Mig.83}.

%%%%%%%%%%%%%%%%%%%%%%%%%%%%%%%%%%%%%%%%%%%%%%%%%%%%%%%%%%%%%%%%%%%%%%%%%%%%%%
\begin{figure*}[ht]
\includegraphics[width=16.0cm]{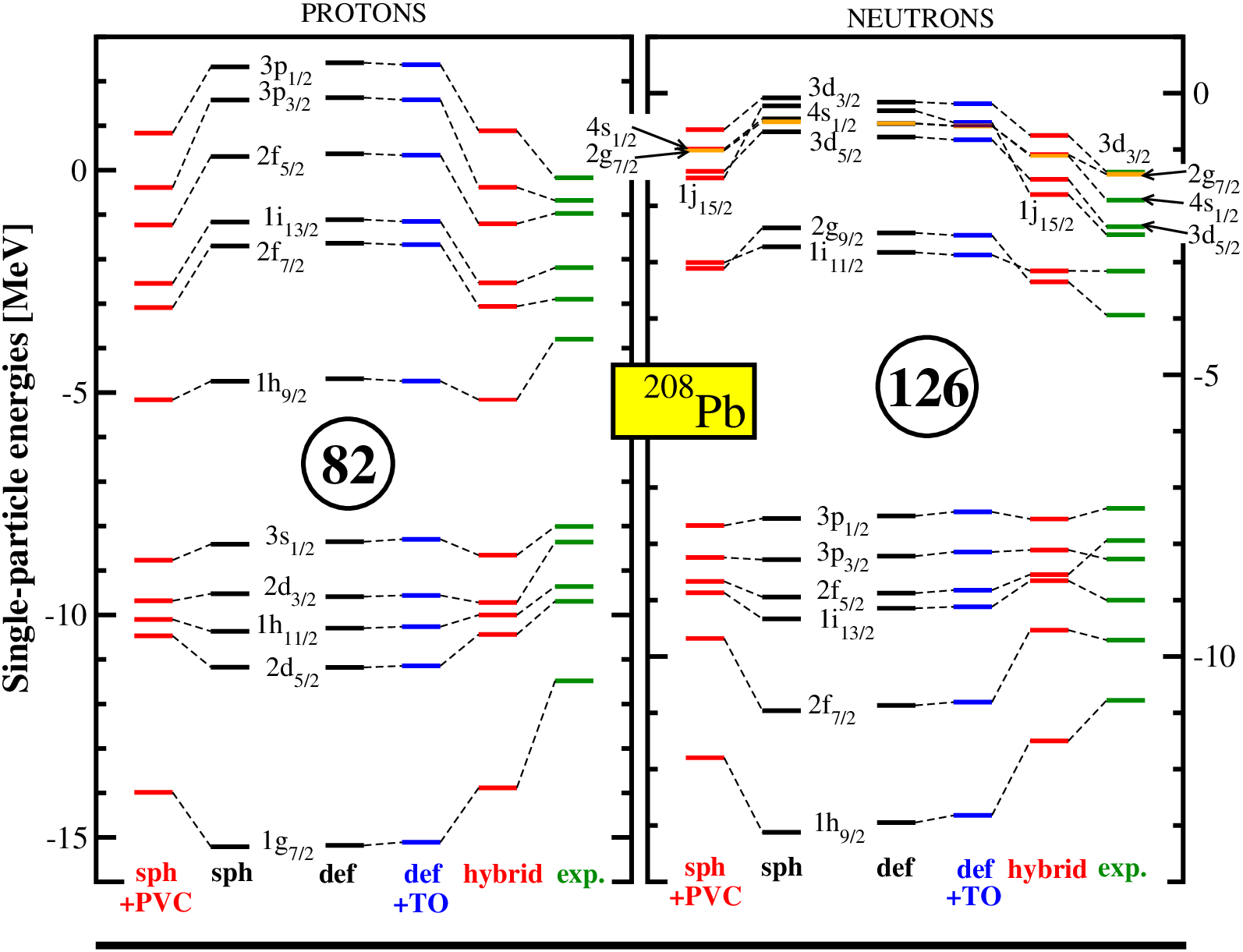}
\caption{(Color online) The same as in Fig.\ \ref{Ni56-fig} but for 
the spectra of $^{208}$Pb. The experimental single-particle levels are based
on the data of Refs.\ \cite{AWT.03,M.91,M.93}.}
\label{Pb208-fig}
\end{figure*}
%%%%%%%%%%%%%%%%%%%%%%%%%%%%%%%%%%%%%%%%%%%%%%%%%%%%%%%%%%%%%%%%%%%%%%%%%%%%%%

  Taking the quasiparticle-phonon coupling into account as an
additional time dependent part of the nucleonic self-energy leads to
the splitting of each mean-field single-particle state into many energy
levels. The dominant levels, i.e. the levels with the largest
spectroscopic factors, shift, as a rule, towards the Fermi level
improving considerably the agreement with data, see the review of
various applications in Ref. \cite{MBBD.85} and references therein.
This result is obtained, however, only for the states in about
10 MeV vicinity of the Fermi energy. For the states lying far from the
Fermi surface one observes a very strong fragmentation making an
extraction of the dominant levels impossible.

  Similar general picture has been obtained within the particle-vibration 
coupling extension of the covariant density functional theory \cite{LR.06}. 
Unlike the non-relativistic versions of the model, in the extended CDFT 
the particle-phonon coupling self-energy allows intermediate nucleonic 
propagation through the Dirac sea states with negative energies, in addition 
to the particle and hole states in the Fermi sea. The contribution of the 
Dirac-sea terms to the self-energy at the Fermi level are, however, found 
negligibly small due to their large energy denominators.

%%%%%%%%%%%%%%%%%%%%%%%%%%%%%%
\subsubsection{Approximations}
%%%%%%%%%%%%%%%%%%%%%%%%%%%%%%

 In the present investigation we retain basically calculational
scheme of Ref. \cite{LR.06}. This approach implies linearized
version of particle-vibration coupling model: the Dyson equation
for single-particle Green function contains phonon-coupling
self-energy with mean-field intermediate nucleonic propagators
and phonon energies and vertices computed within relativistic
random phase approximation (RRPA) \cite{RMGVVC.01}. For medium-mass 
and heavy nuclei the RRPA gives a very reasonable description of the phonon spectra.
Although a more precise description of the phonons can be achieved
within an approach like relativistic time blocking approximation
(RTBA) \cite{LRT.08}, particle-vibration coupling model of Ref.
\cite{LR.06} by using the RRPA phonons takes into account 
the major contribution of the vibrational motion to the 
single-particle spectra.

The application of the model to the lightest doubly magic nuclei
$^{16}$O, $^{40}$Ca, and $^{48}$Ca requires, however, a special
consideration because their mean-field single-particle level
densities are too small to provide a sizable configuration mixing
around the Fermi surface. For $^{16}$O and $^{40}$Ca, it is not
possible, in principle, to reproduce within the RRPA the low-lying 
phonons with positive parity (first
positive parity particle-hole excitations are at too high energies).
In such cases the phonon spectra for the nucleonic self-energy
should be calculated allowing an extension for coupling to more
complex configurations; that implies a generalized particle-vibration 
coupling model. This will be done in a separate work. In $^{48}$Ca 
we have obtained a rather reasonable phonon spectrum, but the strong 
fragmentation of the hole states does not allow extraction of 
dominant single-particle levels.

 The phonons of unnatural parities are known to play only a marginal 
role in the nucleonic self-energy, therefore, they are not included into
the phonon space. Moreover, the inclusion of pairing vibrations into
the phonon space has not been studied yet within the relativistic
framework. In principle, they may bring some additional corrections
to the results, although our approximation is justified by the
results of Refs.\ \cite{LRT.08,LRT.07}, where the fragmentation of the
collective modes has been reproduced very well without inclusion of
the pairing vibrations.

%%%%%%%%%%%%%%%%%%%%%%%%%%%%%%%%%%%%%%%%%%%%%%%%%%%%%%%%%%
\subsection{The numerical details of calculations}
\label{Num-det}
%%%%%%%%%%%%%%%%%%%%%%%%%%%%%%%%%%%%%%%%%%%%%%%%%%%%%%%%%%

  The CRMF equations are solved in the basis of an anisotropic
three-dimensional harmonic oscillator in Cartesian coordinates
characterized by the deformation parameters $\beta_0=0.0$ and
$\gamma=0^{\circ}$ as well as the oscillator frequency $\hbar
\omega_0= 41 A^{-1/3}$ MeV. The truncation of the basis is performed in
such a way that all states belonging to the shells up to fermionic
$N_F$=20 and bosonic $N_B$=20 are taken into account in the
calculations. Numerical analysis indicates that this truncation
scheme provides sufficient numerical accuracy for the physical
quantities of interest. Note that the same truncation of basis is
used in the RRPA and PVC calculations, but they are performed at 
spherical shape.

 The NL3* \cite{NL3*} parametrization of the RMF Lagrangian is 
used in the CRMF, RRPA and PVC calculations. This recently 
fitted parametrization has been successfully applied to the
description of binding energies \cite{NL3*} , ground state 
properties of deformed nuclei \cite{SRH.10}, fission barriers 
\cite{AAR.10}, rotating nuclei \cite{NL3*}, giant resonances 
\cite{NL3*}, and breathing mode \cite{GLLM.10}.

%%%%%%%%%%%%%%%%%%%%%%%%%%%%%%%%%%%%%%%%%%%%%%%%%%%%%%%%%%%%%%
\subsection{Cut-off problem of the phonon basis}
\label{Cut-off}
%%%%%%%%%%%%%%%%%%%%%%%%%%%%%%%%%%%%%%%%%%%%%%%%%%%%%%%%%%%%%%

   Phonons of the multipolarities 2$^+$, 3$^-$, 4$^+$, 5$^-$,
6$^+$ with energies below 15 MeV are included in the model space of
the PVC calculations. We have neglected only the phonon
modes with very small transition probabilities, less than 5\% of
the maximal ones for each J$^{\pi}$. The phonon energies and their
coupling vertices have been computed within the self-consistent
RRPA.

   A rather good convergence of the phonon coupling self-energy
with extension of the phonon space has been obtained numerically:
with very few exceptions, addition of phonon modes with energies
above 15 MeV does not affect the results. Partial contributions from
the phonons with low angular momenta and low energies have been also
investigated. An example of such a study is shown in Fig.
\ref{n1g92} for the neutron 1g$_{9/2}$ state in $^{132}$Sn. This
typical deeply bound hole state is strongly fragmented when the PVC
is taken into account, see the panel (d) with the results obtained
with the full phonon set below 15 MeV.

Panels (a) and (b) show the results obtained when only the phonons
below 5 and 10 MeV, respectively, are taken into account. One can
see that in the former case the picture is very different from the
one displayed in the panel (d). In the latter case the results are
already close to saturation, but there are still two states
between $-20$ and $-18$ MeV with the spectroscopic factors larger than
0.1. Panel (c) shows how the angular momentum cut-off weakens the
fragmentation effect. Similar conclusion can be drawn for other
states either far from or around the Fermi surface: cut-off on the
angular momenta and on the energies of phonons has to be made with
care.

%%%%%%%%%%%%%%%%%%%%%%%%%%%%%%%%%%%%%%
\section{Results and discussion}
\label{Res-disc}
%%%%%%%%%%%%%%%%%%%%%%%%%%%%%%%%%%%%%%

%%%%%%%%%%%%%%%%%%%%%%%%%%%%%%%%%%%%%%%%%%%%%%%%%%
\subsection{Theoretical and experimental spectra}
%%%%%%%%%%%%%%%%%%%%%%%%%%%%%%%%%%%%%%%%%%%%%%%%%%

  An example of the comparison of the results of calculations
and experiment is shown in Fig.\ \ref{Ni56-fig}. Column ``sph''
shows single-particle spectra obtained in spherical RMF calculations 
of even-even $^{56}$Ni nucleus. Column ``def'' displays one-nucleon 
separation energies [defined according to Eqs.\ (\ref{Eq-part}) and 
(\ref{Eq-hole})] obtained in triaxial CRMF calculations in which 
only deformation polarization effects are taken into account. It 
is seen that these energies are  close to the single-particle
energies obtained in spherical calculations. This shows that 
deformation polarization effects induced by extra particle or hole 
are rather modest. In general, they lead to a slight compression 
of the calculated spectra as compared to those obtained in 
spherical calculations. Note that the magnitude and
impact of deformation polarization effects is similar in our
calculations and in the calculations of Ref.\ \cite{RBRMG.98}.

%%%%%%%%%%%%%%%%%%%%%%%%%%%%%%%%%%%%%%%%%%%%%%%
\begin{figure}
\includegraphics[width=8.0cm]{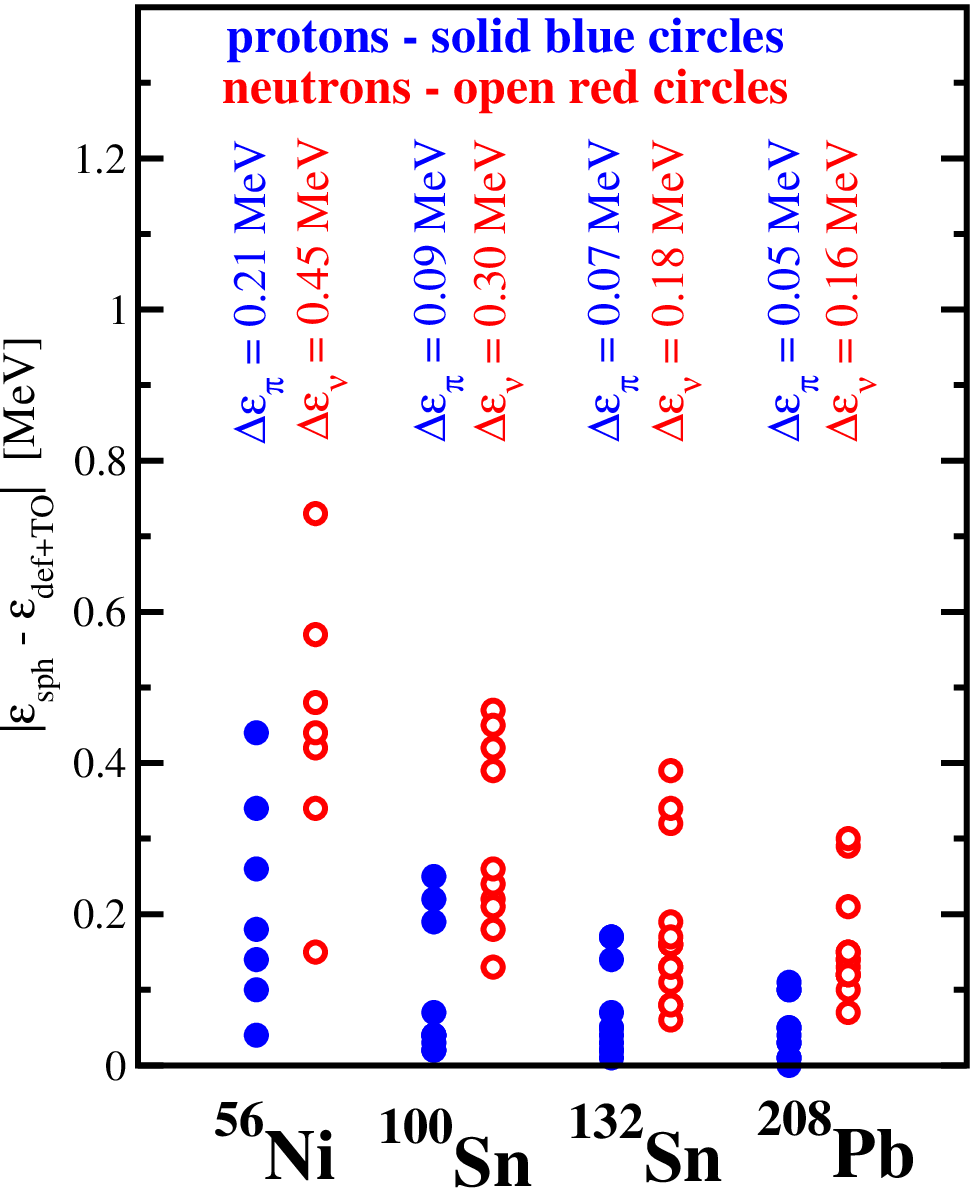}
\caption{ (Color online) Combined polarization effects 
due to deformation and time-odd mean fields. For each
single-particle state of Figs.\ \ref{Ni56-fig}-\ref{Pb208-fig}, 
the absolute values of the difference of the energies obtained 
in spherical (the $\varepsilon_{\rm sph}$ energies; column ``sph''
in Figs.\ \ref{Ni56-fig}-\ref{Pb208-fig}) and deformed (the 
$\varepsilon_{\rm def+TO}$ energies of the calculations with TO 
mean fields included; column ``def+TO'' in Figs.\ 
\ref{Ni56-fig}-\ref{Pb208-fig}) calculations are shown. The 
average polarization effects per state $\Delta \varepsilon_{k}$ 
($k=\pi,\nu$) are also shown.}
\label{Polarization}
\end{figure}
%%%%%%%%%%%%%%%%%%%%%%%%%%%%%%%%%%%%%%%%%%%%%%%

 The inclusion of time-odd mean fields induces additional binding
in odd-mass nuclei (see Ref.\ \cite{AA.10} for more detail) which is
rather modest being around 100-200 keV in the nuclei around $^{56}$Ni.
This again introduces slight compression of the calculated spectra
(Fig.\ \ref{Ni56-fig}). One should note that additional binding
due to TO mean fields in odd-mass nuclei (and a subsequent
compression of the spectra) is substantially larger in the
calculations of Ref.\ \cite{RBRMG.98}. For example, it is around 0.5
MeV in the nuclei around $^{208}$Pb and reaches $1-1.5$ MeV in
$^{56}$Ni. Note that for the latter nucleus additional binding due
to TO mean fields is only $0.1-0.2$ MeV in our CRMF calculations
(see Fig.\ \ref{Ni56-fig}). The reason for this difference is not
clear. However, additional bindings due to TO mean fields in the
nuclei under study obtained in our CRMF calculations are within the
typical ranges obtained in the systematic study of the impact of TO
mean fields on the binding energies of odd mass nuclei (Ref.\
\cite{AA.10}). In addition, the CRMF code employed here has been
extensively and successfully used in the description of rotating
systems in which TO fields have very large impact on the moments of
inertia (see Refs.\ \cite{VRAL.05,AA.10b} and references quoted
therein). These two facts strongly suggest that the impact of TO
mean fields is correctly described in the current work.

   The dominant levels, i.\ e. the levels with the largest 
spectroscopic factors, as obtained in spherical PVC calculations, 
are shown in column ``sph+PVC''. One can see two effects of 
particle-vibration coupling: (i) the general compression of 
the spectra leading to a better agreement with experiment and 
(ii) in some cases the change of the level sequences. With few 
exceptions, the spectroscopic factors of the dominant levels in 
the vicinity of the Fermi level vary between 0.5 and 0.9 (see 
Tables \ref{Table-spec-1} and \ref{Table-spec-2} below), thus 
these states retain basically their single-particle nature.

  The impact of particle-vibration coupling on specific state
can be quantified by the energy difference between the energies 
of this state in the columns ``sph'' and ``sph+PVC'' of Fig.\ 
\ref{Pb208-fig}. These energy differences treated as the corrections 
due to PVC are then added to one-neutron separation energies obtained 
in triaxial CRMF calculations; this leads to column ``hybrid''. 
The results in this column are compared with experimental data 
presented in column ``exp''.

  The single-particle spectra of doubly magic $^{100}$Sn, $^{132}$Sn 
and $^{208}$Pb nuclei, obtained in the same way as Fig.\ 
\ref{Ni56-fig}, are displayed in Figs.\ \ref{Sn100-fig}, \ref{Sn132-fig} 
and \ref{Pb208-fig}. They show similar effects due to PVC and the 
polarizations induced by deformation and TO  mean fields as in the 
case of $^{56}$Ni. Their details will be discussed below.

  Let consider major conclusions emerging from these calculations. 
They are related to the role of polarizations effects due to deformation 
and TO mean fields and the impact of PVC on the accuracy of the 
description of single-particle spectra, shell gaps, pseudospin doublets 
and spin-orbit splittings.

%%%%%%%%%%%%%%%%%%%%%%%%%%%%%%%%%%%%%%%%%%%%%%%%%%%%%
\subsection{Polarization effects due to deformation 
and time-odd mean fields}
\label{Pol-eff}
%%%%%%%%%%%%%%%%%%%%%%%%%%%%%%%%%%%%%%%%%%%%%%%%%%%%%

  Combined polarization effects due to deformation and TO mean 
fields are shown in Fig.\ \ref{Polarization}. Two features 
are clearly seen.

  First, combined polarization effects due to deformation 
and time-odd mean fields decrease with an increase of mass 
number. Indeed, with an increase of the size of nucleus the 
relative role of each single-particle orbital becomes smaller 
which leads to the fact that the calculated quadrupole 
deformations of odd-mass nuclei decrease with an increase of 
mass. For example, the average proton and neutron deformations 
of calculated single-particle states in $^{57}$Cu and $^{57}$Ni 
are 0.045 and 0.060, respectively. On the contrary, these 
deformations are significantly smaller in $^{209}$Bi and 
$^{209}$Sn; they are only 0.0055 and 0.006, respectively.
Thus, odd nuclei neighboring to $^{208}$Pb remain basically
spherical. The decrease of calculated deformations of 
odd mass nuclei neighboring to doubly magic nuclei with
increasing mass clearly indicates that deformation 
polarization effects also decrease with mass number. In 
addition, additional binding due to TO mean fields 
decreases with mass (see Ref.\ \cite{AA.10} for details).
The combination of these two effects is responsible for
the features seen in Fig.\ \ref{Polarization}.

%%%%%%%%%%%%%%%%%%%%%%%%%%%%%%%%%%%%%%%%%%%%%%%%%%%%%%%%%%%%%%%%%%%%%%%%%%%%%%
\begin{figure}
\includegraphics[width=8.0cm]{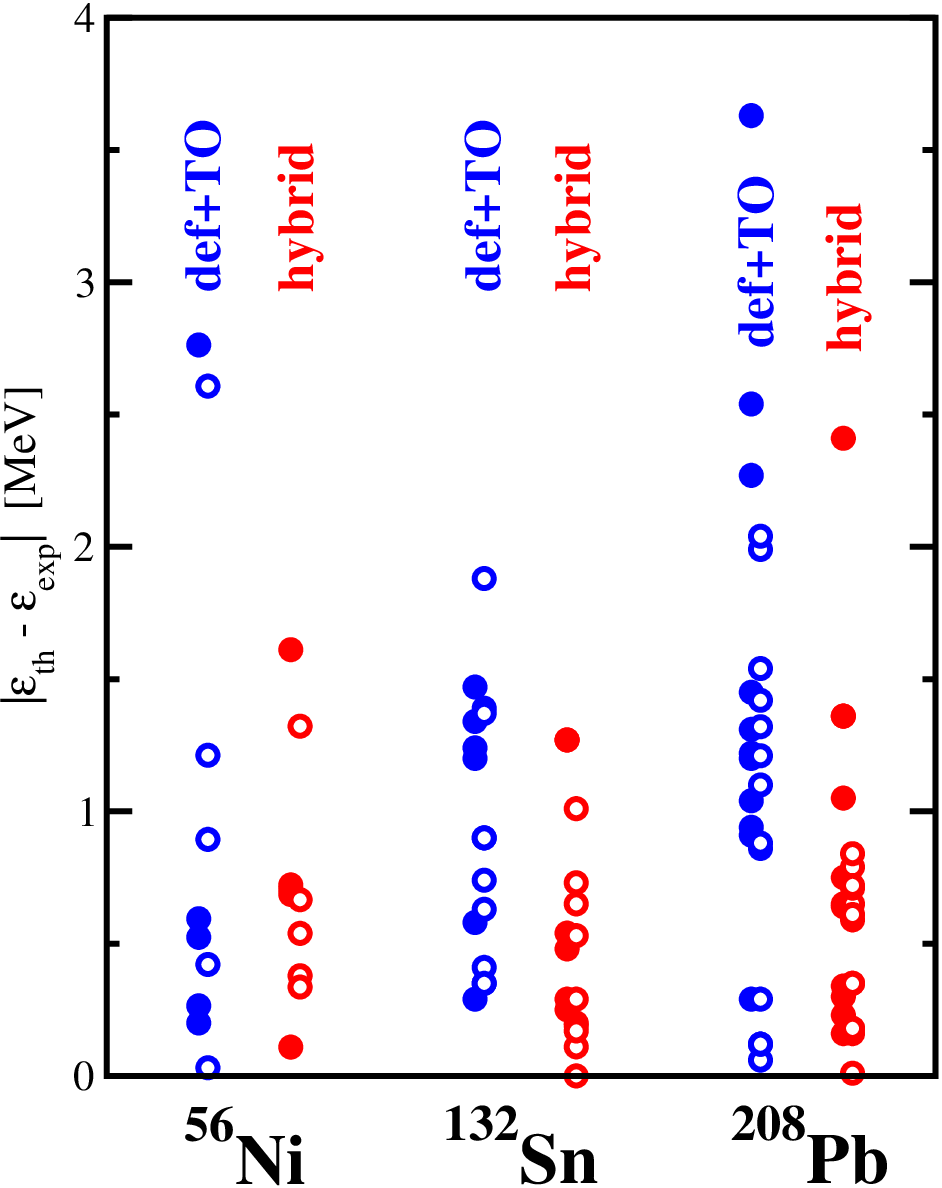}
\caption{(Color online) The deviations of calculated energies of the 
single-particle states from experimental ones obtained in the ``def+TO'' 
and ``hybrid'' calculational schemes for indicated nuclei. The 
results for proton and neutron states are given by solid and open 
circles.}
\label{Dev-fig}
\end{figure}
%%%%%%%%%%%%%%%%%%%%%%%%%%%%%%%%%%%%%%%%%%%%%%%%%%%%%%%%%%%%%%%%%%%%%%%%%%%%%%

 Fig.\ \ref{Polarization} also reveals that combined 
polarization effects due to deformation and TO mean 
fields are smaller for protons than for neutrons by 
a factor of approximately two. This is especially true
for the $N=Z$ $^{56}$Ni and $^{100}$Sn nuclei which have 
the same proton and neutron single-particle states. 
The part of this difference comes from the fact 
that additional binding due to TO mean fields is smaller 
for odd-proton nuclei as compared with odd-neutron ones
(see Ref.\ \cite{AA.10}). The analysis of Ref.\ \cite{AA.10} 
clearly indicates that the contributions of the Coulomb 
force to the proton single-particle energies in the 
presence of TO mean fields is responsible for this 
feature. Deformation polarization effects show the same 
features, namely, they are smaller for odd-proton nuclei as 
compared with odd-neutron  ones. For example, average deformation 
polarizations per single-particle state are 0.157 MeV 
and 0.326 MeV for proton and neutron single-particle 
states of $^{56}$Ni and 0.060 MeV and 0.164 MeV for 
proton and neutron single-particle states of $^{100}$Sn.
Similar to the case of polarizations due to TO mean 
fields, the origin of this difference has to be traced 
back to the Coulomb force since compared sets of 
proton and neutron single-particle states are the 
same in a given $N=Z$ nucleus.

The majority of the comparisons of the energies of the
single-particle states obtained in different particle
vibration coupling models with experiment neglect 
deformation and TO mean field polarization effects 
induced by the odd particle (see, for example, Refs.\ 
\cite{MBBD.85,LR.06,CSFB.10}). The current analysis 
indicates that within the framework of CDFT this neglect 
is more or less justified only for heavy nuclei, and it 
is more justified for proton subsystem than for neutron 
one.

%%%%%%%%%%%%%%%%%%%%%%%%%%%%%%%%%%%%%%%%%%%%%%%%%%%%%%%%%%
\subsection{The accuracy of the description of 
                  single-particle spectra}
\label{SP-acc}
%%%%%%%%%%%%%%%%%%%%%%%%%%%%%%%%%%%%%%%%%%%%%%%%%%%%%%%%%%

   The analysis of Figs.\ \ref{Ni56-fig}, \ref{Sn132-fig},
and \ref{Pb208-fig} reveals that the inclusion of particle
vibration coupling substantially improves the accuracy of
the description of single-particle spectra. This statement
is quantified in Fig.\ \ref{Dev-fig} and Table 
\ref{Table-dev}.

  Fig.\ \ref{Dev-fig} displays the distribution of the deviations 
between calculated and experimental energies of the single-particle 
states in proton and neutron subsystems of $^{56}$Ni, $^{132}$Sn 
and $^{209}$Pb. It clearly shows that {\it in average} the 
inclusion of particle-vibration coupling substantially improves 
the description of single-particle spectra, and that neutron 
single-particle states are better described than proton ones. However, 
even in the ``hybrid'' calculations there are few (mostly proton) 
states  which deviate from experiment by more than 1 MeV. Note that 
the inclusion of particle-vibration coupling  can make the agreement 
between theory and experiment worse for some states. This is seen, for 
example, in the case of proton $3s_{1/2}$ and $1h_{9/2}$ subshells in 
$^{208}$Pb (Fig.\ \ref{Pb208-fig}).

%%%%%%%%%%%%%%%%%%%%%%%%%%%%%%%%%%%%%%%%%%%%%%%%%%%%%%%%%%%%%%%%%%%%%
\begin{table}[h]
\caption{Average deviations per state $\Delta\varepsilon$ between 
calculated and experimental energies of the single-particle 
states for a proton (neutron) subsystem of a given nucleus. The 
results obtained in the ``def+TO'' and ``hybrid'' calculational 
schemes  are shown.}
\begin{center}
\begin{tabular}{|c|c|c|} \hline
Nucleus/subsystem  &  $\Delta\varepsilon_{def+TO}$ [MeV] & $\Delta\varepsilon_{hybrid}$ [MeV] \\ 
\hline
 $^{56}$Ni/proton    &   0.76    &  0.77    \\
 $^{56}$Ni/neutron   &   0.89    &  0.71   \\
 $^{132}$Sn/proton   &   1.02    &  0.68   \\
 $^{132}$Sn/neutron  &   0.89    &  0.39   \\
 $^{208}$Pb/proton   &   1.53    &  0.84    \\
 $^{208}$Pb/neutron  &   1.00    &  0.47    \\ \hline
\end{tabular}
\end{center}
\label{Table-dev}
\end{table}
%%%%%%%%%%%%%%%%%%%%%%%%%%%%%%%%%%%%%%%%%%%%%%%%%%%%%%%%%%%%%%%%%%%%%%%%%%%%%%%%%%%%%%%%%%

%%%%%%%%%%%%%%%%%%%%%%%%%%%%%%%%%%%%%%%%%%%%%%%%%%%%%%%%%%%%%%%%%%%%%%%%%%%%%%
\begin{figure}
\includegraphics[width=8.0cm]{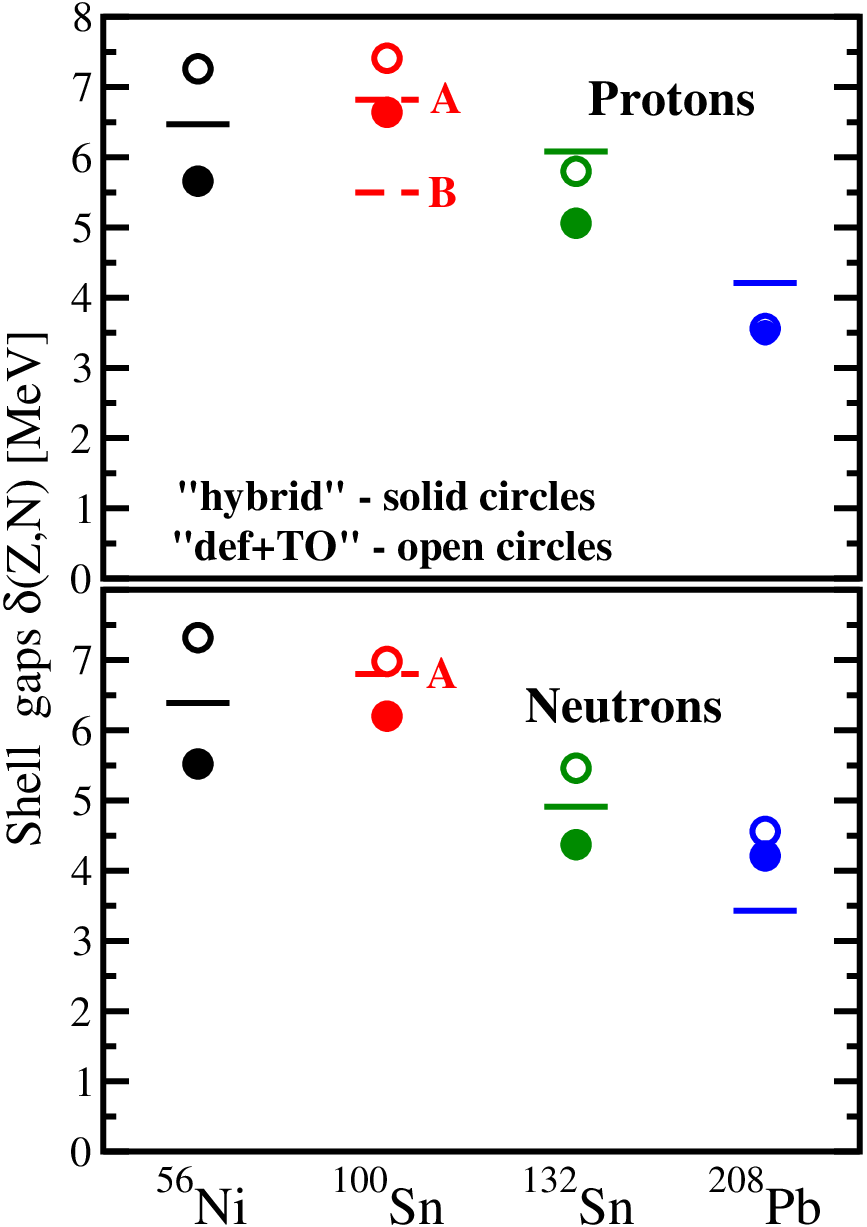}
\caption{(Color online) The size of proton and neutron shell gaps.
The results of the ``def+TO'' and ``hybrid'' calculations are
compared with experimental values shown by horizontal lines. In the case 
of $^{100}$Sn, the experimental shell gaps are 
extracted from extrapolated single-particle energies of Refs.\ 
\cite{Ext-1} (A) and \cite{Ext-2} (B) [see also Table \ref{Table-sn100}]. 
Note that the size of the $Z=82$ shell gap in $^{208}$Pb is almost the 
same in both calculations.}
\label{Gaps-fig}
\end{figure}
%%%%%%%%%%%%%%%%%%%%%%%%%%%%%%%%%%%%%%%%%%%%%%%%%%%%%%%%%%%%%%%%%%%%%%%%%%%%%%

  Average deviations per state $\Delta\varepsilon$ between calculated 
and experimental energies of the single-particle states are shown 
in Table \ref{Table-dev}. They are defined as 
\begin{equation}
\Delta \varepsilon = \frac{\sum_{i=1}^N |\varepsilon_i^{th}-\varepsilon_i^{exp}|}{N}
\end{equation}
where $N$ is the number of the states with known experimental 
single-particle energies, and $\varepsilon_i^{th}$ ($\varepsilon_i^{exp}$) 
are calculated (experimental) energies of the single-particle states. 
One can see that the inclusion of PVC substantially improves the description 
of the single-particle states in $^{132}$Sn and $^{208}$Pb. On the contrary, 
PVC introduces no (small) improvement in the description of the proton 
(neutron) single-particle states of $^{56}$Ni. It is interesting to mention
that the best agreement with experiment is obtained in the ``hybrid''
calculations of neutron-rich $^{132}$Sn.

   While the accuracy of the description of proton and neutron states is 
comparable in $^{56}$Ni, neutron states are appreciable better described 
than proton ones in the ``hybrid'' calculations of $^{132}$Sn and 
$^{208}$Pb. The analysis of Figs.\ \ref{Sn132-fig} and \ref{Pb208-fig} 
strongly suggests that the increases of the depth of proton potential in 
these two nuclei by few hundreds keV (and as a result the lowering of the 
energies of proton single-particle states by similar energy) will improve 
the agreement with experiment. This clearly indicates that the inclusion 
of the single-particle energies into the fit of the CDFT parametrizations 
can provide an extra information on the depth of the proton and neutron 
potentials and their evolution with particle numbers.

%%%%%%%%%%%%%%%%%%%%%%%%%%%%%%%%%%%%%%%%%%%%%%%%%%%%%%%%%%%%%%%%%%%%
\begin{table}[h]
\caption{Extrapolated ``experimental'' single-particle energies
of spherical proton and neutron subshells in $^{100}$Sn. The 
results of the extrapolations of Refs.\ \cite{Ext-1,Ext-2} are shown.}
\begin{center}
\begin{tabular}{|c|c|c|} \hline
  Subshell         &  $\varepsilon_i$(MeV) \cite{Ext-1}  & $\varepsilon_i$(MeV) \cite{Ext-2} \\ \hline
\multicolumn{3}{|c|}{  Neutrons }            \\ \hline 
    $1h_{11/2}$    &  -8.6(5)                  &  -7.8(8)    \\
    $2d_{3/2}$     &  -9.2(5)                  &  -8.8(8)    \\
    $3s_{1/2}$     &  -9.3(5)                  &  -9.3(9)    \\
    $2g_{7/2}$     &  -10.93(20)               &  -10.4(10)  \\
    $2d_{5/2}$     &  -11.13(20)               &  -11.1(10) \\
    $1g_{9/2}$     &  -17.93(30)               &             \\
    $2p_{1/2}$     &  -18.38(20)               &             \\ \hline
%------------------------------------------------------------
\multicolumn{3}{|c|}{ Protons }            \\ \hline 
%------------------------------------------------------------
    $1g_{7/2}$     &  +3.90(15)                &   2.6(3)    \\
    $2d_{5/2}$     &  +3.00(80)                &   2.8(3)    \\
    $1g_{9/2}$     &  -2.92(20)                &  -2.9(3)    \\
    $2p_{1/2}$     &  -3.53(20)                &  -3.5(3)    \\
    $2p_{3/2}$     &  -6.38                    &             \\
    $1f_{5/2}$     &  -8.71                    &             \\ \hline
\end{tabular}
\end{center}
%\vspace{0.5cm}
\label{Table-sn100}
\end{table}
%%%%%%%%%%%%%%%%%%%%%%%%%%%%%%%%%%%%%%%%%%%%%%%%%%%%%%%%%%%%%%%%%%%%%%%

 The detailed comparison between theory and experiment
has not been performed for $^{100}$Sn since
very little experimental information is available on neighboring
to $^{100}$Sn nuclei with one nucleon less (more) (see, for example,
Refs.\ \cite{Sn101-1,Sn101-2}). Unfortunately, the extrapolations of 
the single-particle energies from other odd nuclei towards $^{100}$Sn do 
not bring unique values. This is illustrated in Table \ref{Table-sn100} 
where the results of the extrapolations of Refs.\ \cite{Ext-1,Ext-2} 
are shown. Note that the estimates of the energies of the single-particle
states of Ref.\ \cite{Ext-2} coincide within the estimation errors
(given in Table \ref{Table-sn100} in parentheses) with the values
reported in Ref.\ \cite{Ext-1}, except for the energy of the $1g_{7/2}$
proton state. However, estimation errors are significant which 
makes detailed comparison between theory and experiment meaningless. On 
the other hand, one can clearly see in Fig.\ \ref{Sn100-fig} that 
PVC improves the agreement with experimental estimates.

%%%%%%%%%%%%%%%%%%%%%%%%%%%%%%%%%%%%%%%%%%%%%%%%%%%%%%%%%%%%%%%%%%%%
\begin{table*}[ptb]
\caption{ Spectroscopic factors $S$ of the dominant
single-particle levels in odd nuclei surrounding $^{208}$Pb and
$^{132}$Sn calculated within relativistic particle-vibration
coupling model compared to experimental data. The experimental data
are taken from Refs.\ \cite{EV.68} ($^{209}$Bi, ($^3$He,d)
reaction), \cite{209Bi} ($^{209}$Bi, the ($\alpha$,t) reaction),
\cite{207Tl} ($^{207}$Tl, the (d,$^{3}$He) reaction), \cite{BFIA.70}
($^{207}$Tl, the (t,$\alpha$) reaction), \cite{MCD.70} ($^{207}$Pb,
the (d,t) reaction), \cite{GWL.80} ($^{207}$Pb, the
($^3$He,$\alpha$) reaction), \cite{EKV.69} ($^{209}$Pb, the (d,p)
reaction), \cite{TG.75} ($^{209}$Pb, the ($\alpha$, $^{3}$He)
reaction) and \cite{Sn133} ($^{133}$Sn, the (d,p) reaction). For comparison
the experimental data from two different reactions are presented for
odd mass nuclei neighboring to $^{208}$Pb. }
\begin{center}
\vspace{1mm}
\tabcolsep=1.40em \renewcommand{\arraystretch}{0.9}%
\begin{tabular}
[c]{ccccc|cccc}\hline\hline \
Nucleus & State & $S_{th}$ & $S_{exp}$ & $S_{exp}$ & Nucleus & State & $S_{th}$ & $S_{exp}$\\
\hline
& & & & & & & \\
$^{209}$Pb & 2g$_{9/2}$ & 0.85 & 0.78$\pm$0.1 \cite{EKV.69}& 0.94 \cite{TG.75}  & $^{133}$Sn & 2f$_{7/2}$ & 0.89 & 0.86$\pm$0.16 \\
\  & 1i$_{11/2}$ &  0.89 & 0.96$\pm$0.2 \cite{EKV.69} & 1.05 \cite{TG.75}       &            & 3p$_{3/2}$ & 0.91 & 0.92$\pm$0.18 \\
\  & 1j$_{15/2}$ &  0.66 & 0.53$\pm$0.2 \cite{EKV.69} & 0.57 \cite{TG.75}       &            & 1h$_{9/2}$ & 0.88 & \\
\  & 3d$_{5/2}$ &  0.89 & 0.88$\pm$0.1 \cite{EKV.69} &                 &            & 3p$_{1/2}$ & 0.91 & 1.1$\pm$0.3 \\
\  & 4s$_{1/2}$ &  0.92 & 0.88$\pm$0.1 \cite{EKV.69} &                 &            & 2f$_{5/2}$ & 0.89 & 1.1$\pm$0.2 \\
\  & 2g$_{7/2}$ &  0.87 & 0.78$\pm$0.1 \cite{EKV.69} &                 &            & & & \\
\  & 3d$_{3/2}$ &  0.89 & 0.88$\pm$0.1 \cite{EKV.69} &                 &            & & & \\
&  &  &  & & & & & \\
$^{209}$Bi & 1h$_{9/2}$ &  0.88 & 1.17 \cite{EV.68}      & 0.80 \cite{209Bi}& $^{133}$Sb & 1g$_{7/2}$ & 0.86 & \\
\  & 2f$_{7/2}$ & 0.78  & 0.78 \cite{EV.68}      &  0.76 \cite{209Bi}      &            & 2d$_{5/2}$ & 0.82 & \\
\  & 1i$_{13/2}$ & 0.63 & 0.56 \cite{EV.68}      & 0.74 \cite{209Bi}       &            & 2d$_{3/2}$ & 0.63 & \\
\  & 2f$_{5/2}$ &  0.61 & 0.88 \cite{EV.68}      & 0.57 \cite{209Bi}       &            & 1h$_{11/2}$ & 0.79 & \\
\  & 3p$_{3/2}$ &  0.62 & 0.67 \cite{EV.68}      & 0.44 \cite{209Bi}       &            & 3s$_{1/2}$ & 0.70& \\
\  & 3p$_{1/2}$ &  0.37 & 0.49 \cite{EV.68}      & 0.20 \cite{209Bi}       &            & & & \\
&  &  &   & & & & & \\
$^{207}$Pb & 3p$_{1/2}$ & 0.90 & &1.08 \cite{GWL.80} & $^{131}$Sn  & 2d$_{3/2}$ & 0.88 & \\
\  & 2f$_{5/2}$ &  0.87 & 1.13 \cite{MCD.70} &1.05 \cite{GWL.80}       &             & 1h$_{11/2}$ & 0.86 & \\
\  & 3p$_{3/2}$ &  0.86 & 1.00 \cite{MCD.70} &0.95 \cite{GWL.80}       &             & 3s$_{1/2}$ & 0.87 & \\
\  & 1i$_{13/2}$ & 0.82 & 1.04 \cite{MCD.70} &0.61 \cite{GWL.80}       &             & 2d$_{5/2}$ & 0.70 & \\
\  & 2f$_{7/2}$ &  0.64 & 0.89 \cite{MCD.70} &0.64 \cite{GWL.80}       &             & 1g$_{7/2}$ & 0.72 & \\
\  & 1h$_{9/2}$ &  0.38 &  &    &            &       &   & \\
&  &  &  & & & & & \\
$^{207}$Tl & 3s$_{1/2}$ &  0.84 & 0.95  \cite{BFIA.70} & 0.85 \cite{207Tl} & $^{131}$In  & 1g$_{9/2}$ & 0.85 & \\
\  & 2d$_{3/2}$ &  0.86 & 1.15 \cite{BFIA.70} & 0.90 \cite{207Tl}         &             & 2p$_{3/2}$& 0.70 & \\
\  & 1h$_{11/2}$ &  0.80 & 0.89 \cite{BFIA.70}  & 0.88 \cite{207Tl}        &             & 2p$_{1/2}$ & 0.85 & \\
\  & 2d$_{5/2}$ &  0.68 & 0.62 \cite{BFIA.70} & 0.63 \cite{207Tl}         &             & 1f$_{5/2}$ & 0.37 & \\
\  & 1g$_{7/2}$ & 0.22 & 0.40 \cite{BFIA.70} & 0.27 \cite{207Tl} &
& & & \\\hline\hline \label{pbsn}
\end{tabular}
\end{center}
\label{Table-spec-1}
\end{table*}
%%%%%%%%%%%%%%%%%%%%%%%%%%%%%%%%%%%%%%%%%%%%%%%%%%%%%%%%%%%%%%%%%%%

%%%%%%%%%%%%%%%%%%%%%%%%%%%%%%%%%%%%%%%%%%%%%%%%%%%%%%%%%%%%%%%%%%%%%
\subsection{The impact of particle-vibration coupling on shell gaps}
%%%%%%%%%%%%%%%%%%%%%%%%%%%%%%%%%%%%%%%%%%%%%%%%%%%%%%%%%%%%%%%%%%%%%

  The size of the shell gap (between last occupied and first unoccupied
states in even-even doubly magic nucleus) is defined as
\begin{equation}
\delta(Z,N)=min\{(\varepsilon_{i})^{above}\}-max\{(\varepsilon_{i})^{below}\}
\end{equation}
where $(\varepsilon_{i})^{above}$ and $(\varepsilon_{i})^{below}$ stand for
the energies of the single-particle states above and below the shell
gap. This definition of the shell gap incorporates the information
from three different nuclei because the single-particle energies are
defined according to Eqs.\ (\ref{Eq-part}) and (\ref{Eq-hole}).  
The advantage of this definition is that it allows to incorporate the 
polarization (due to deformation and TO mean fields) and PVC effects
on the size of shell gap. 

 Fig.\ \ref{Gaps-fig} compares experimental shell gaps with the shell 
gaps obtained in the ``def+TO'' and ``hybrid'' calculations. The 
extrapolations of Refs.\ \cite{Ext-1,Ext-2} for the single-particle 
energies in $^{100}$Sn lead to quite different values of the proton $Z=50$ 
shell gap. It is reasonable to expect the same situation for the neutron 
$N=50$ shell gap, the size of which, however, cannot be determined in the 
extrapolations of Ref.\ \cite{Ext-2} since no estimates for the 
single-particle states below it are provided (see Table \ref{Table-sn100}). 

 The shell gaps are largest in the spherical mean field calculations. 
The gaps become smaller with the inclusion of each additional type of
correlations. The inclusion of polarization effects due to deformation
and TO mean fields decreases slightly their sizes (see Figs.\ 
\ref{Ni56-fig}, \ref{Sn100-fig}, \ref{Sn132-fig} and \ref{Pb208-fig}).
Thus, we compare with experiment in Fig.\ \ref{Gaps-fig} only the PVC 
(``hybrid'') and best mean field (``def+TO'') calculations.
This figure shows that particle-vibration coupling 
decreases substantially the size of the shell gaps. The effect is 
most pronounced in $^{56}$Ni where PVC decreases the proton $Z=28$ 
and neutron $N=28$ shell gaps by almost 2 MeV. On the contrary, the 
effect of particle-vibration coupling is least pronounced in 
$^{208}$Pb in which it decreases the size of the neutron $N=126$ 
gap by only few hundreds keV and has almost no impact on the proton 
$Z=82$ shell gap.

  Comparing experimental and calculated shell gaps in $^{56}$Ni, 
$^{132}$Sn and $^{208}$Pb, one can conclude that both calculational 
schemes provide similar accuracy of the description of experimental 
data. However, experimental shell gaps are typically overestimated 
in the ``def+TO'' calculations. The exceptions are the $Z=50$ gap in 
$^{132}$Sn and $Z=82$ gap in $^{208}$Pb. On the contrary, with 
exception of the $N=126$ gap in $^{208}$Pb, the size of experimental 
gaps is underestimated in the calculations with particle-vibration 
coupling.

%%%%%%%%%%%%%%%%%%%%%%%%%%%%%%%%%%%%%%%%%%%%%%%
\subsection{Spectroscopic factors}
\label{Spec-fac}
%%%%%%%%%%%%%%%%%%%%%%%%%%%%%%%%%%%%%%%%%%%%%%%%

%%%%%%%%%%%%%%%%%%%%%%%%%%%%%%%%%%%%%%%%%%%%%%%%%%%%%%%%%%%%%%%%%%%
\begin{table*}[ptb]
\caption{Spectroscopic factors $S$ of the dominant
single-particle levels in odd nuclei surrounding $^{100}$Sn and
$^{56}$Ni calculated within relativistic particle-vibration coupling
model. They are compared to the experimental values extracted by means
of the $(d,p)$ transfer \cite{LTLHS.09} and one-neutron knockout 
\cite{Ni56-knock} reactions in the case of $^{57}$Ni.}%
\label{snni}
\begin{center}
\vspace{1mm}
\tabcolsep=1.40em \renewcommand{\arraystretch}{0.8}%
\begin{tabular}
[c]{cccc|ccccc}\hline\hline \
Nucleus & State & $S_{th}$ &\ & Nucleus & State & 
$S_{th}$ & $S_{exp}$ \cite{LTLHS.09} & $S_{exp}$ \cite{Ni56-knock} \\
\hline
& & & & & & & & \\
$^{101}$Sn & 2d$_{5/2}$ & 0.85 & & $^{57}$Ni & 2p$_{3/2}$& 0.83 & 0.95$\pm$0.29 
& 0.58$\pm$0.11  \\
\  & 1g$_{7/2}$ & 0.85 & & & 1f$_{5/2}$ & 0.79 & 1.40$\pm$0.42 & \\
\  & 2d$_{3/2}$ & 0.78 & & & 2p$_{1/2}$ & 0.76 & 1.00$\pm$0.30 & \\
\  & 3s$_{1/2}$ & 0.81 & & & 1g$_{9/2}$ & 0.79 & & \\
\  & 1h$_{11/2}$& 0.80 & & & & & & \\
&  &  &  & & & & & \\
$^{101}$Sb & 2d$_{5/2}$ & 0.87 & & $^{57}$Cu & 2p$_{3/2}$ & 0.85 & & \\
\  & 1g$_{7/2}$ & 0.86 & & & 1f$_{5/2}$ & 0.80 & & \\
\  & 2d$_{3/2}$ & 0.83 & & & 2p$_{1/2}$ & 0.80 & & \\
\  & 3s$_{1/2}$ & 0.87 & & & 1g$_{9/2}$ & 0.80 & & \\
\  & 1h$_{11/2}$& 0.81 & & & & & & \\
&  &  &   & & & & & \\
$^{99}$Sn & 1g$_{9/2}$ & 0.84 & & $^{55}$Ni  & 1f$_{7/2}$ & 0.78 & & \\
\  & 2p$_{1/2}$ & 0.85 & & & 2s$_{1/2}$ & 0.71 & & \\
\  & 2p$_{3/2}$ & 0.71 & & & 1d$_{3/2}$ & 0.62 & & \\
\  & 1f$_{5/2}$ & 0.62 & & & 1d$_{5/2}$ & 0.20 & & \\
\  & 1f$_{7/2}$ & 0.11 & & & & & & \\
&  &  &  & & & & & \\
$^{99}$In & 1g$_{9/2}$ &  0.85 & & $^{55}$Co  & 1f$_{7/2}$ & 0.78 & & \\
\  & 2p$_{1/2}$ & 0.86 & & & 2s$_{1/2}$ & 0.73 & & \\
\  & 2p$_{3/2}$ & 0.73 & & & 1d$_{3/2}$ & 0.64 & & \\
\  & 1f$_{5/2}$ & 0.66 & & & 1d$_{5/2}$ & 0.20 & & \\
\  & 1f$_{7/2}$ & 0.14 & & & & & \\
 \hline\hline
\end{tabular}
\end{center}
\label{Table-spec-2}
\end{table*}
%%%%%%%%%%%%%%%%%%%%%%%%%%%%%%%%%%%%%%%%%%%%%%%%%%%%%%%%%%

%%%%%%%%%%%%%%%%%%%%%%%%%%%%%%%%%%%%%%%%%%%%%%%%%%%%%%%%%%%%%%%%%%%%%%%%%%%%%%
\begin{figure}
\includegraphics[width=8.0cm]{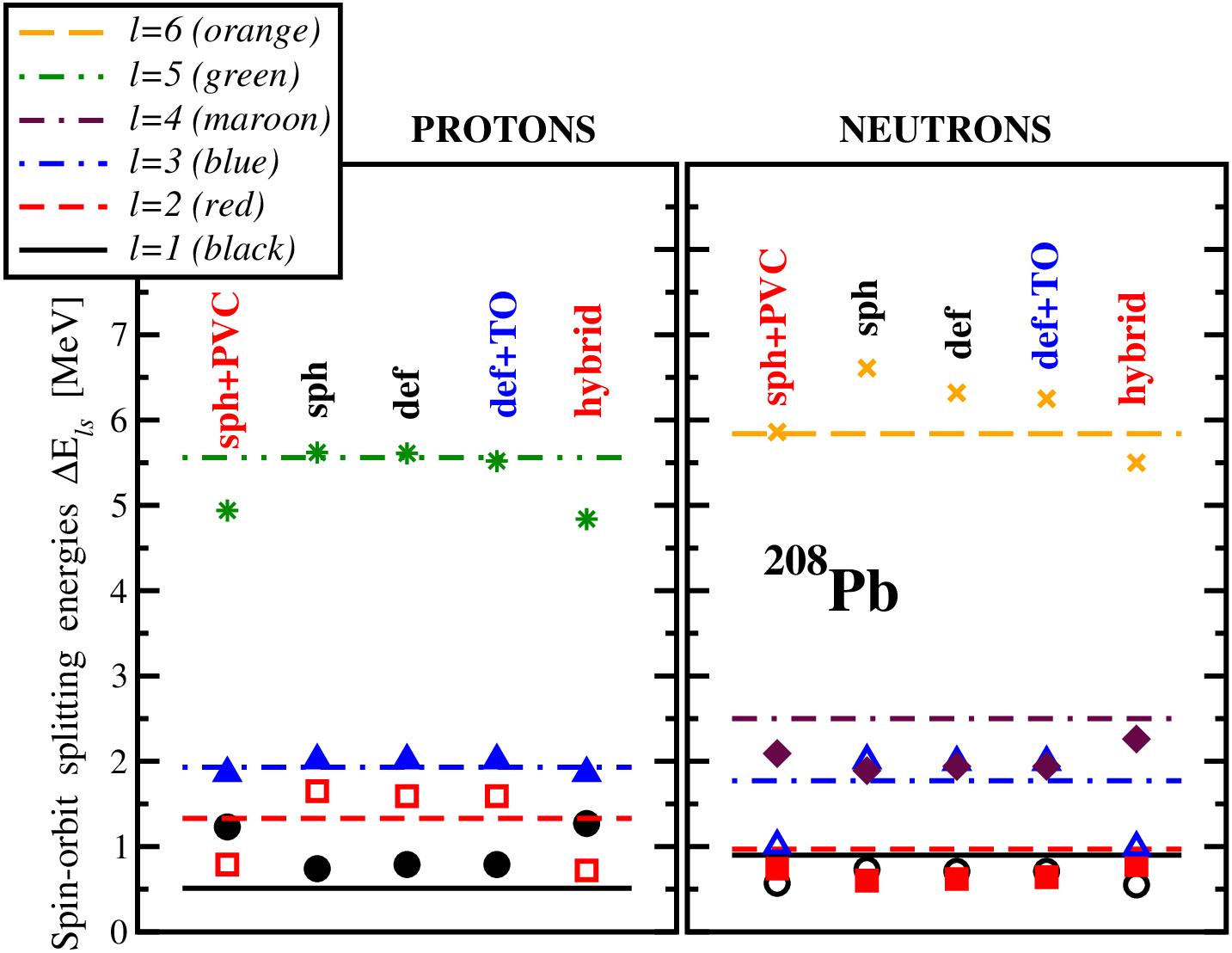}
\caption{(Color online) Spin-orbit splitting energies $\Delta E_{ls}$ of the 
spin-orbit doublets in $^{208}$Pb. Insert shows the colors used for different 
values of orbital angular momentum $l$. Theoretical results are shown by symbols,
while the experimental values by solid lines. Solid (open) symbols are
used for spin-orbit doublets which are build from the particle (hole) 
states with respect of the $^{208}$Pb core. The stars and crosses are used for the
spin-orbit doublets which involve both particle and hole states.}
\label{SO-pb208}
\end{figure}
%%%%%%%%%%%%%%%%%%%%%%%%%%%%%%%%%%%%%%%%%%%%%%%%%%%%%%%%%%%%%%%%%%%%%%%%%%%%%%

The calculated spectroscopic factors for odd nuclei with doubly-magic cores 
are compared with experiment in Tables \ref{pbsn} and \ref{snni}. In odd 
nuclei neighboring to $^{208}$Pb, the experimental spectroscopic factors 
are reasonably well reproduced in the PVC calculations. 
%Only for the state 
%$1h_{9/2}$ in $^{207}$Pb, there is no good agreement between the PVC calculations
%and experiment. 
Both in experiment and in calculations the proton states 
are found to be somewhat more fragmented than the neutron ones. When comparing 
theory with experiment one should keep in mind that the experimental spectroscopic 
factors depend considerably on the parameters used in the model 
analysis and on employed reaction, which is clearly
seen in the staggering of experimental values. Note that the results obtained 
for $^{208}$Pb with the NL3* parametrization are very close to those obtained 
previously in Ref.\ \cite{LR.06} with the NL3 force.

  For other nuclei, the experimental data are available only for $^{133}$Sn 
(Table \ref{pbsn}) and $^{57}$Ni (Table \ref{snni}). The level of agreement 
between theory and experiment in $^{133}$Sn is comparable with the one seen in 
odd nuclei neighbouring to
$^{208}$Pb. The situation in $^{57}$Ni is more controversial. 
 For example, the sum rule (\ref{sum_rule}) is strongly violated in the case of
the $1f_{5/2}$ state in experimental data of Ref.\ \cite{Ni56-knock}. In addition,
experimental spectroscopic factors for the $2p_{3/2}$ state in $^{57}$Ni extracted by means
of the $(d,p)$ transfer \cite{LTLHS.09} and one-neutron knockout \cite{Ni56-knock} 
reactions differ by a factor of almost two. Currently there is no satisfactory 
explanation for this difference (Ref.\ \cite{LTLHS.09}). The results of the
PVC calculations are lower than the experimental values obtained in the $(d,p)$ 
transfer reaction but higher than the ones obtained in the one-neutron knockout
reaction. On the other hand, the spectroscopic factors obtained in the PVC
calculations are very close to the ones obtained in the spherical shell model
calculations in full $fp$ model space which employ GXPF1A interaction (see Table
II in Ref.\ \cite{LTLHS.09}). Note that the latter calculations provide best 
agreement with the data in the Ni isotopes \cite{LTLHS.09}. One should also 
mention that non-relativistic particle-vibration calculations of Ref.\ 
\cite{B.09} give the spectroscopic factors for the $2p_{3/2}$, $1f_{5/2}$,
and $2p_{1/2}$ states which are very close to 0.6.

%%%%%%%%%%%%%%%%%%%%%%%%%%%%%%%%%%%%%%%%%%%%%%%%%%%%%%%%%%
\subsection{The impact of particle-vibrational coupling
on the spin-orbit splittings}
%%%%%%%%%%%%%%%%%%%%%%%%%%%%%%%%%%%%%%%%%%%%%%%%%%%%%%%%%%

  Covariant density functional theory naturally describes the 
spin-orbit interaction in nuclei \cite{Rei.89,Rin.96}, which 
is a relativistic effect that has to be added phenomenologically 
in non-relativistic models. The comparison of experimental 
spin-orbit splittings with calculations has been performed 
only on the mean field level so far (see, for example, Refs.\ 
\cite{RBRMG.98,BRRMG.99}). However, the current work clearly
indicates that particle-vibration coupling has an impact on 
the energies of the single-particle states. Thus, it is 
reasonable to expect that it will also modify the spin-orbit 
splittings.

%%%%%%%%%%%%%%%%%%%%%%%%%%%%%%%%%%%%%%%%%%%%%%%%%%%%%%%%%%%%%%%%%%%%%%%%%%%%%%
\begin{figure}
\includegraphics[width=8.0cm]{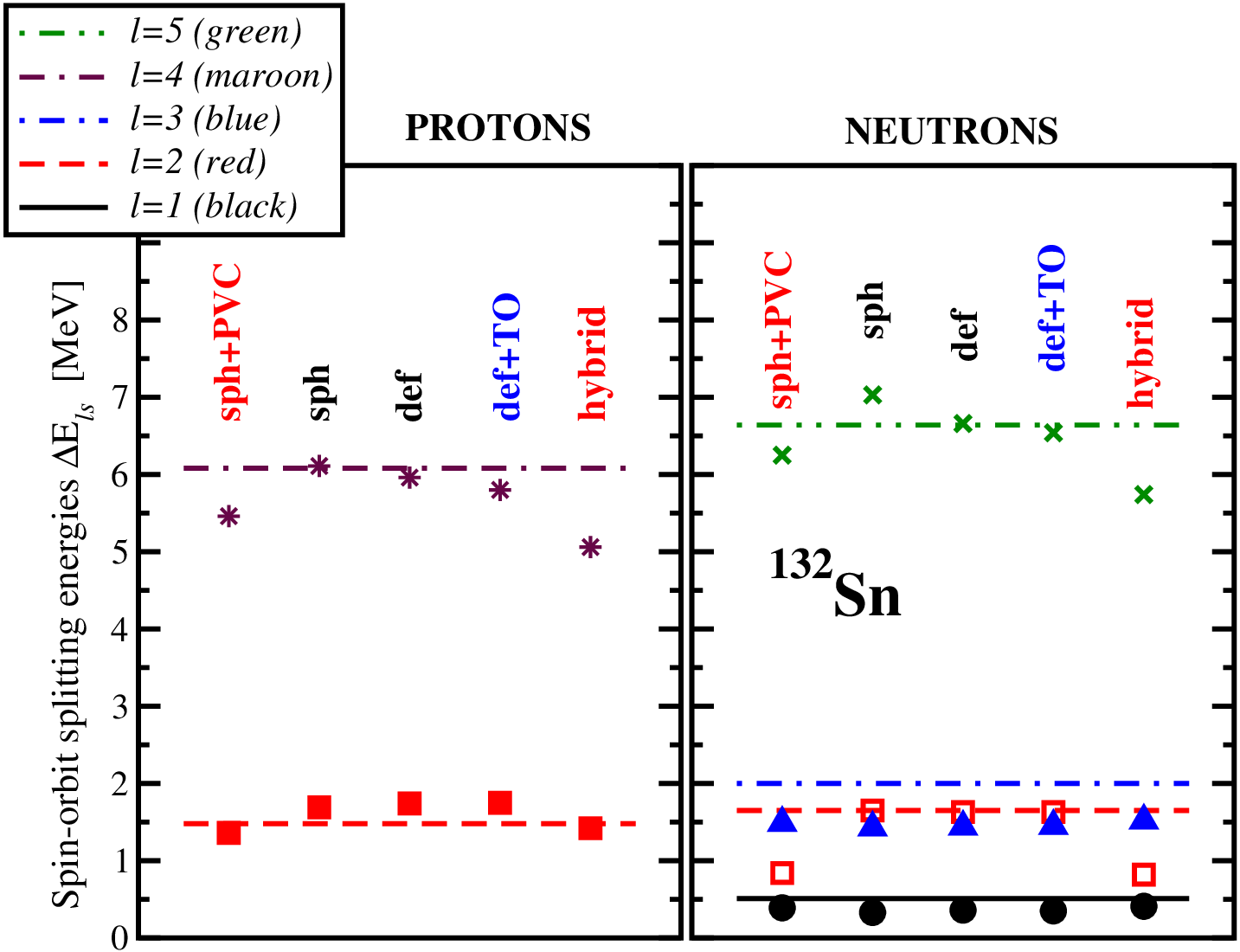}
\caption{(Color online) The same as in Fig.\ \ref{SO-pb208} but for the 
spin-orbit doublets of $^{132}$Sn.}
\label{SO-sn132}
\end{figure}
%%%%%%%%%%%%%%%%%%%%%%%%%%%%%%%%%%%%%%%%%%%%%%%%%%%%%%%%%%%%%%%%%%%%%%%%%%%%%%

\begin{figure}
\includegraphics[width=8.0cm]{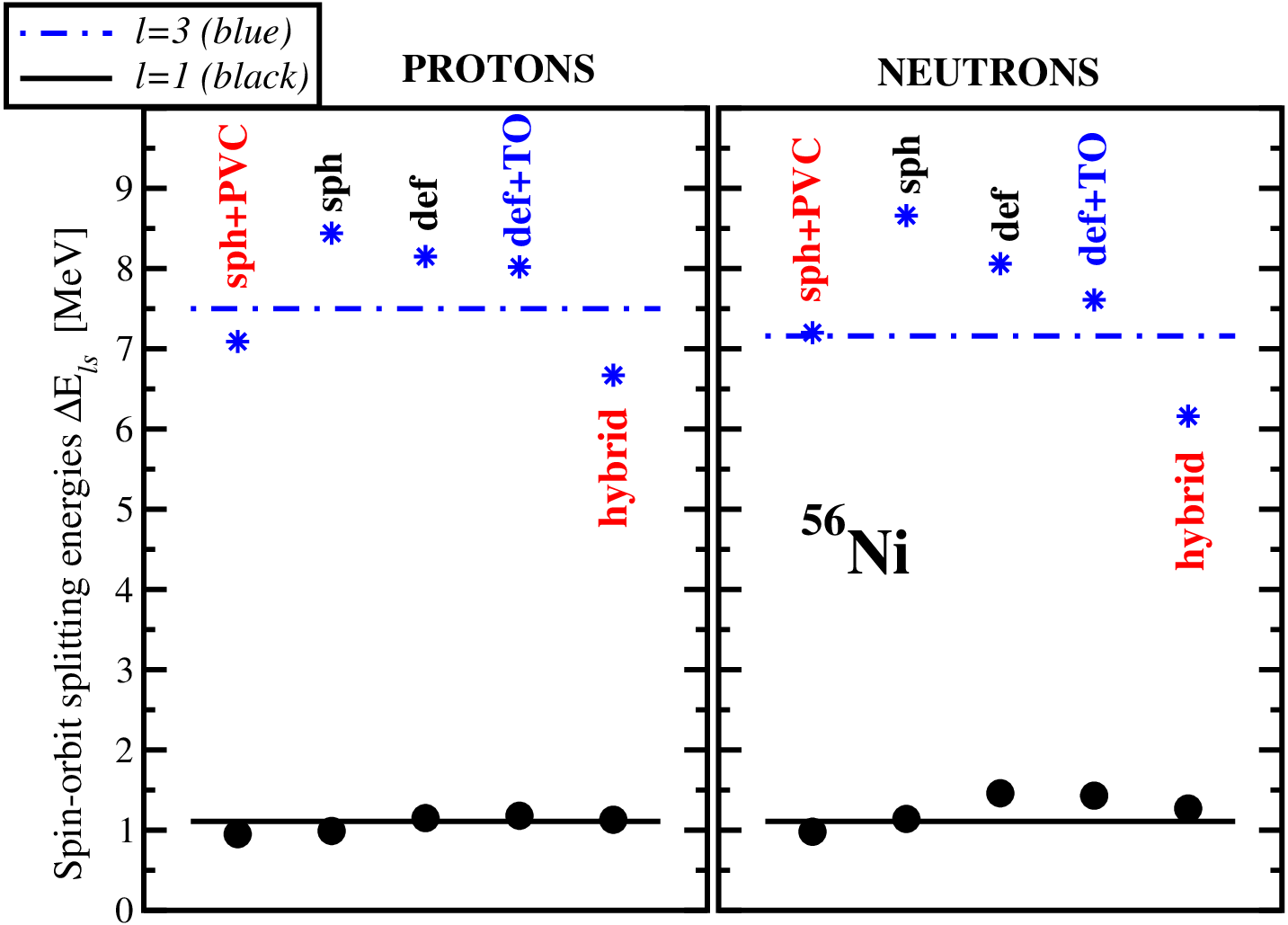}
\caption{(Color online) The same as in Fig.\ \ref{SO-pb208} 
but for the spin-orbit doublets of $^{56}$Ni.}
\label{SO-ni56}
\end{figure}

  In order to understand the impact of polarization effects
(due to deformation and TO mean fields) and particle-vibration
coupling on spin-orbit splittings, the results of the 
``sph'', ``sph+PVC'', ``def'', ``def+TO'', and ``hybrid'' 
calculations are compared with available experimental data on the
spin-orbit doublets in Figs.\ \ref{SO-pb208}, \ref{SO-sn132} 
and \ref{SO-ni56}.
 
  The impact of deformation and TO mean field polarization effects 
on spin-orbit splittings is rather modest in $^{208}$Pb (Fig.\ 
\ref{SO-pb208}) since these effects are smallest in heavy systems 
(see Sec.\ \ref{Pol-eff}). In addition, both of them lead to a 
compression of the single-particle spectra. As a consequence, they
act in the same direction for the members of the spin-orbit 
doublet which is built either from hole (below the shell gap) or 
particle (above the shell gap) states. Moreover, these polarization
effects are more or less similar in magnitude for the members of 
such spin-orbit doublets. As a consequence, they cancel to a large 
extent when spin-orbit splitting is calculated as the difference of 
the energies of the members of the spin-orbit doublet. Such behavior 
is typical for the spin-orbit doublets based on the orbitals with low 
orbital angular momentum {\it l}, both members of which are located
either below or above the shell gap. For this type of the spin-orbit 
doublets, the impact of polarization effects on spin-orbit splittings 
is also small in $^{132}$Sn (Fig.\ \ref{SO-sn132}). However, it becomes 
more pronounced (especially, in neutron subsystem) in $^{56}$Ni 
(Fig.\ \ref{SO-ni56}).

  The situation is different for the spin-orbit doublets built on
the orbitals with high orbital angular momentum {\it l}. For such
doublets, one member is located above the shell gap, while another
below the shell gap. The compression of the single-particle spectra
due to deformation and TO mean field polarization effects leads to 
the decrease of spin-orbit splitting. This decrease is especially
pronounced for light nuclei and for neutron subsystem in a given
nucleus (compare Figs.\ \ref{SO-ni56}, \ref{SO-sn132}, and 
\ref{SO-pb208}) because of particle number and subsystem (proton or 
neutron) dependencies of polarization effects discussed in Sec.\ 
\ref{Pol-eff}.

  The impact of particle-vibration coupling on spin-orbit splittings
is state-dependent. It is small or modest in the proton $\it l=3$ 
$(2f_{7/2}-2f_{5/2})$ and neutron $\it l=1$ $(3p_{3/2}-3p_{1/2})$, 
$\it l=2$ $(2d_{5/2}-2d_{3/2})$, and $\it l=4$ $(2g_{9/2}-2g_{7/2})$ 
doublets of $^{208}$Pb (Fig.\ 
\ref{SO-pb208}), proton $\it l=2$ $(2d_{5/2}-2d_{3/2})$ and neutron 
$\it l=1$ $(3p_{3/2}-3p_{1/2})$, $\it l=3$ $(2f_{7/2}-2f_{5/2})$ doublets 
of $^{132}$Sn (Fig.\ \ref{SO-sn132}) and neutron and proton $\it l=1$ 
$(2p_{3/2}-2p_{1/2})$ doublets in $^{56}$Ni (Fig.\ \ref{SO-ni56}).
In these doublets, the corrections  to the energies of the 
members of the spin-orbit doublet due to PVC are more or less the same.

 However, this is not always the case for the doublets based on the 
orbitals with low orbital angular momentum {\it l}. For example, the
corrections due to PVC differ significantly for the proton $3d_{3/2}$
and $3d_{5/2}$ states of $^{208}$Pb (Fig.\ \ref{Pb208-fig}). As a 
consequence, particle-vibration coupling has significant effect on the
spin-orbit splitting of the proton $\it l=2$ $(3d_{5/2}-3d_{3/2})$
doublet. In addition, spin-orbit splittings of the proton  
$\it l=1$ $(3p_{3/2}-3p_{1/2})$ and neutron $\it l=3$ $(2f_{7/2}-2f_{5/2})$
doublets in $^{208}$Pb (Fig.\ \ref{SO-pb208}) as well as of neutron 
$\it l=2$ $(2d_{5/2}-2d_{3/2})$ doublet in $^{132}$Sn (Fig.\ \ref{SO-sn132})
are strongly affected by particle-vibration coupling.

  The effect of particle-vibration coupling is especially pronounced
for the spin-orbit doublets with high orbital angular momentum $l$, the 
one member of which is located below the shell gap and another above 
the shell gap. These are the proton $\it l=5$ $(1h_{11/2}-1h_{9/2})$ 
and neutron $\it l=6$ $(1i_{13/2}-1i_{11/2})$ doublets in $^{208}$Pb
(Fig.\ \ref{SO-pb208}), the proton $\it l=4$ $(1g_{9/2}-1g_{7/2})$ and neutron
$\it l=5$ $(1h_{11/2}-1h_{9/2})$ doublets in $^{132}$Sn (Fig.\ \ref{SO-sn132})
and the proton and neutron $\it l=3$ $(1f_{7/2}-1f_{5/2})$ doublets
in $^{56}$Ni (Fig.\ \ref{SO-ni56}).  For these doublets, the spin-orbit
splittings are largest in spherical calculations.  The inclusion 
of polarization effects due to deformation and TO mean fields decreases 
these splittings.  The PVC coupling further reduces the energy splittings 
by approximately 1 MeV in medium and heavy mass nuclei (Figs.\ \ref{SO-pb208} 
and \ref{SO-sn132}) and by approximately 1.5 MeV in $^{56}$Ni (Fig.\ 
\ref{SO-ni56}).

\begin{figure}
\includegraphics[width=8.0cm]{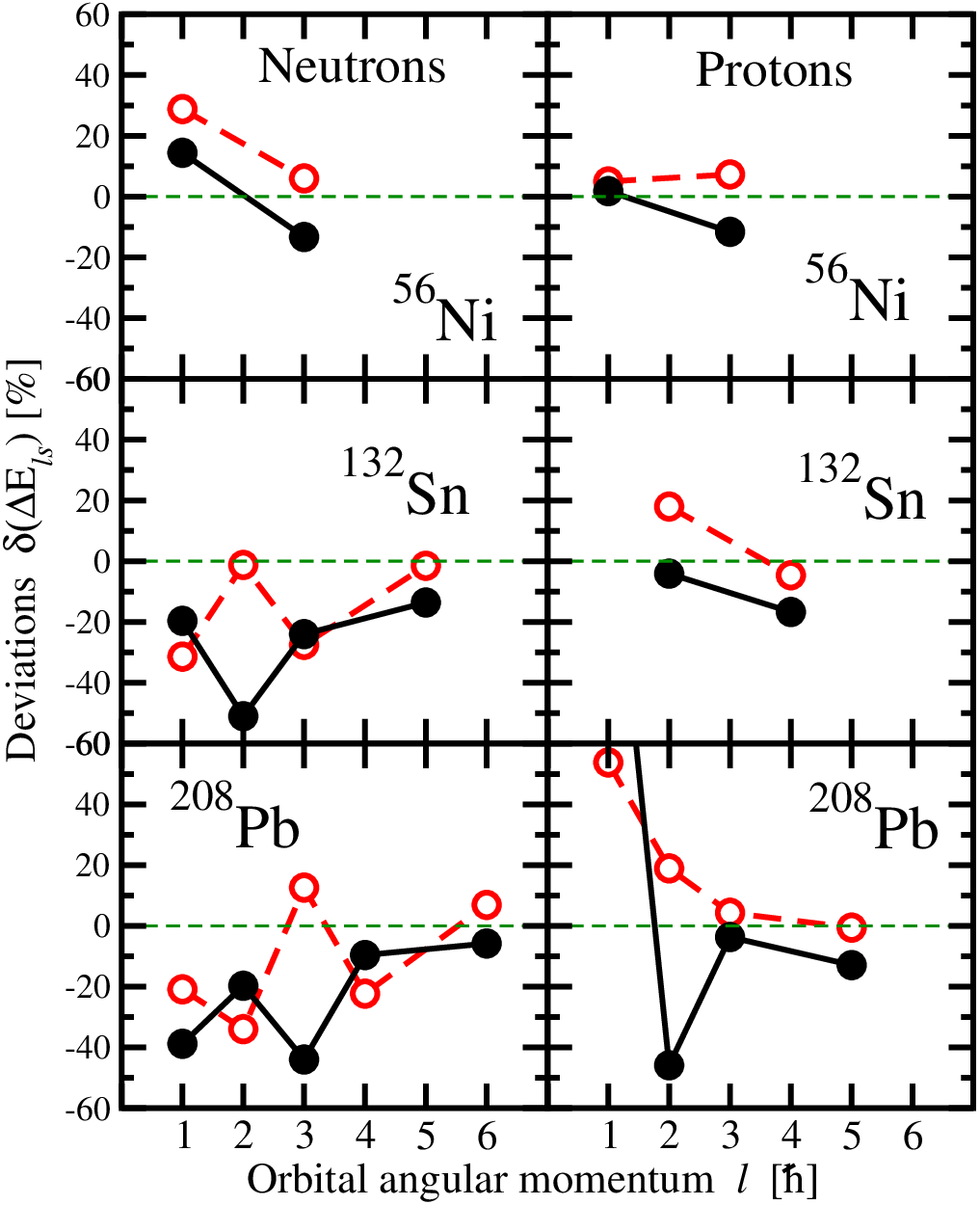}
\caption{ (Color online) Percentual deviations of the spin-orbit splittings 
in the ``def+TO'' (red open circles connected by dashed lines) and ``hybrid'' 
(black solid circles connected by solid lines) from the experimental values.}
\label{SO-dev}
\end{figure}

  Fig. \ref{SO-dev} presents percentual deviations $\delta(\Delta E_{ls})$ 
of the calculated spin-orbit splittings from experimental ones defined as
\begin{eqnarray}
\delta(\Delta E_{ls})=
\frac{\Delta E_{ls}^{th}-\Delta E_{ls}^{exp}}
{\Delta E_{ls}^{exp}} \times 100\%
\end{eqnarray} 
where $\Delta E_{ls}$ is the spin-orbit splitting energy. Negative (positive)
values of $\delta(\Delta E_{ls})$ indicate that experimental spin-orbit 
splitting is underestimated (overestimated) in the calculations. One can see 
that in average the mean-field results are more or less evenly scattered around 
zero percentual deviation. On the contrary, with few exceptions particle-vibration 
coupling calculations underestimate experimental spin-orbit splittings. 
Spin-orbit doublets built on the orbitals with high orbital momentum $l$, 
involving the members across the shell gap, are rather well (within the 20\% 
deviation) described in both calculations. These doublets are characterized by 
large spin-orbit splittings $\Delta E_{ls}$ in the range of 5-7 MeV, so the 20\%
deviation means that the absolute deviation from experiment is typically less 
than 1 MeV. The largest percentual deviations are observed for the spin-orbit 
doublets built on the orbitals with low orbital angular momentum $l$. These 
doublets are characterized by low spin-orbit splittings so relatively low absolute 
deviations from experimental values  of the order of few hundreds keV result
in appreciable percentual deviations.

  The analysis of the results of calculations allows to conclude that 
the inclusion of particle-vibration coupling decreases the accuracy of the 
description of spin-orbit splittings. This is clearly visible when absolute
and/or percentual deviations per doublet are compared in the mean field and PVC
calculations. The absolute deviations per doublet are 0.34 MeV [0.50 MeV],
0.23 MeV [0.56 MeV] and 0.26 MeV [0.45 MeV] in the mean field  (``def+TO'')
[particle-vibration coupling (``hybrid'')] calculations in $^{56}$Ni, 
$^{132}$Sn and $^{208}$Pb, respectively. The percentual deviations per doublet 
are 11.8 \% [10.3\%], 14\% [21.5\%] and 19.3\% [36.4\%] in the mean field 
(``def+TO'') [particle-vibration coupling (``hybrid'')] calculations in 
$^{56}$Ni, $^{132}$Sn and $^{208}$Pb, respectively.

  These results are not surprising and should not be viewed negatively. 
First of all, as discussed in Secs.\ \ref{Exp-vs-th} and \ref{Spec-fac},
the experimental levels are not pure single-particle levels, and as such
they are closer in nature to the calculated levels of the PVC model. 
Second, relativistic description \cite{Rei.89} predicts a definite connection 
between the Dirac effective mass $m^*_D/m$ of the nucleon in the kinetic 
energy and the strength of the spin-orbit force because the same effective 
mass appears in both terms. In addition, the spin-orbit term is sensitive 
to the spatial variations of the effective mass.  Particle-vibration coupling 
affects the effective mass of the nucleon and as a result it has an impact 
on calculated spin-orbit splittings. Third, these results point to a new 
direction of improving the covariant energy density functionals. This is because
empirical energy spacings between spin-orbit partner states in finite nuclei
determine a relatively narrow interval of allowed values for the Dirac effective 
mass $0.57 \leq m^*_D/m \leq 0.61$ on the mean field level \cite{NVR.08}. The 
observed impact of particle-vibration coupling on spin-orbit splittings suggests 
that this interval of the Dirac effective mass $m^*_D/m$ may in reality be 
broader.

%%%%%%%%%%%%%%%%%%%%%%%%%%%%%%%%%%%%%%%%%%%%%%%%%%%%%%%%%%%%%%%%%%%%%%%%%%%%%%
%%%%%%%%%%%%%%%%%%%%%%%%%%%%%%%%%%%%%%%%%%%%%%%%%%%%%%%%%%%%%%%%%%%%%%%%%%%%%%

%%%%%%%%%%%%%%%%%%%%%%%%%%%%%%%%%%%%%%%%%%%%%%%%%%%%%%%%%%%%%%%%%%%%%%%%%%%%%%
%%%%%%%%%%%%%%%%%%%%%%%%%%%%%%%%%%%%%%%%%%%%%%%%%%%%%%%%%%%%%%%%%%%%%%%%%%%%%%

%%%%%%%%%%%%%%%%%%%%%%%%%%%%%%%%%%%%%%%%%%%%%%%%%%%%%%%%%%
\subsection{The impact of particle-vibtational coupling
on the pseudospin doublets}
%%%%%%%%%%%%%%%%%%%%%%%%%%%%%%%%%%%%%%%%%%%%%%%%%%%%%%%%%%

%%%%%%%%%%%%%%%%%%%%%%%%%%%%%%%%%%%%%%%%%%%%%%%%%%%%%%%%%%%%%%%%%%%%%%%%%%%%%%
\begin{figure*}
\includegraphics[width=14.0cm]{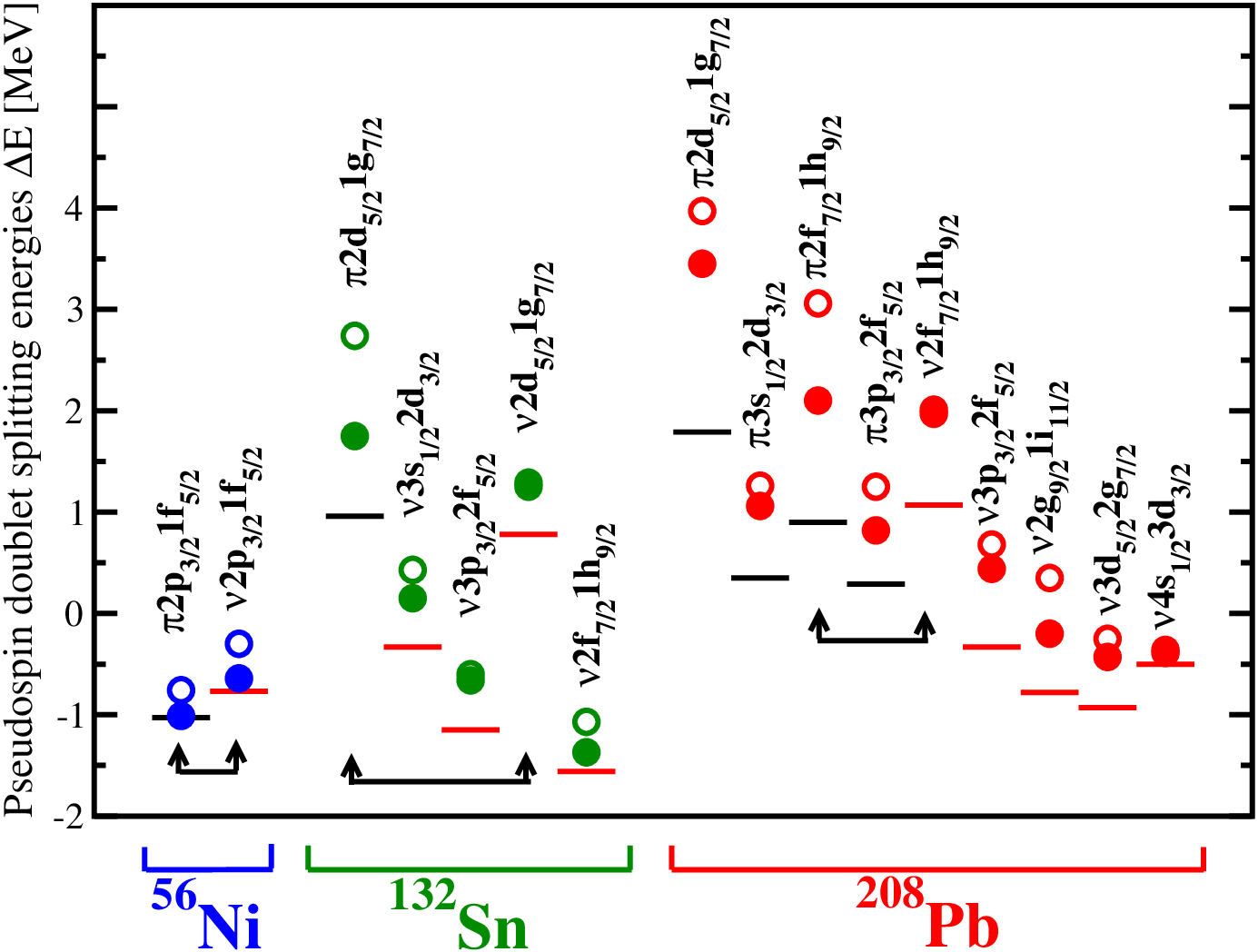}
\caption{(Color online) Pseudospin doublet splitting energies in $^{56}$Ni,
$^{132}$Sn and $^{208}$Pb. Experimental data are shown by horizontal lines;
black lines are for protons, red for neutrons. The results of the
``def+TO'' and ``hybrid'' calculations are shown by open and solid
circles, respectively. If both results are close in energy only solid 
circle is visible. Brackets with arrows indicate proton and neutron
pseudospin doublets with the same single-particle structure. The doublets
are labelled by the spherical subshell labels of their members.}
\label{pseudo}
\end{figure*}
%%%%%%%%%%%%%%%%%%%%%%%%%%%%%%%%%%%%%%%%%%%%%%%%%%%%%%%%%%%%%%%%%%%%%%%%%%%%%%

  It was shown in Ref.\ \cite{Gin.97} that quasidegenerate pseudospin
doublets in nuclei, discovered more than 40 years ago \cite{HA.69,AHS.69},
arise from the near equality in magnitude of attractive scalar
$S$ and repulsive vector $V$ relativistic mean fields in which the
nucleons move\footnote{Extensive review of the manifestations of pseudospin
symmetry in different physical systems is presented in Ref.\ \cite{G.05}.}.
Pseudospin doublets have nonrelativistic quantum
numbers $(n_r, l, j=l+1/2)$ and $(n_r-1, l+2, j=l+1/2)$ where
$n_r$, $l$ and $j$ are the single-nucleon radial, orbital and
total angular momentum quantum numbers, respectively\footnote{Note
that in the current manuscript the single-particle subshells are 
labelled by the labels which use $(n_r+1)$ in the first position 
of the label.}.  

  The pseudospin doublet can be characterized by pseudospin doublet 
splitting energy which is defined as the difference of the energies 
of pseudospin doublet members
\begin{eqnarray}
\Delta E = E_{n_r, l, j=l+1/2} - E_{(n_r-1),(l+2),j=l+3/2}.  
\end{eqnarray}
With this definition of $\Delta E$, the low-$l$ (high-$l$) member 
of the pseudospin doublet is higher in energy than its high-$l$ 
(low-$l$) counterpart if $\Delta E >0$ ($\Delta E < 0$). Note that
the calculations fail to reproduce the relative order (in energy) 
of the members of pseudospin doublet if the signs of $\Delta E$ 
are different in the calculations and experiment.

  Fig.\ \ref{pseudo} compares available experimental data on pseudospin
doublet splitting energies in $^{56}$Ni, $^{132}$Sn and $^{208}$Pb 
with PVC (``hybrid'') and best mean field (``def+TO'')
calculations. Note that the $\Delta E$ values obtained in the  
``sph'' and ``def'' calculations deviate from the ones obtained in 
the ``def+TO'' calculations by less than 200 keV (in the majority 
of the cases the difference is less than 100 keV). This is because
both members of pseudospin doublet are located either above or
below the shell gaps. The polarization effects due to deformation and
TO mean fields act in the same direction for these states. As a 
result, these effects cancel each other to a large degree when the 
difference of the energies of the members of pseudospin doublet is 
taken.

  Fig.\ \ref{pseudo} clearly shows that particle-vibration coupling 
substantially improves the description of splitting energies in pseudospin 
doublets; the average deviations from experiment for $\Delta E$  are 
0.93 MeV and 0.6 MeV in the mean field and PVC calculations, respectively.
There are still some doublets in which the energy splitting $\Delta E$
is poorly reproduced in model calculations. These are $\pi 2d_{5/2} 1g_{7/2}$
and $\pi 2f_{7/2} 1h_{9/2}$ doublets in $^{208}$Pb (see Figs.\ \ref{Pb208-fig} 
and \ref{pseudo}), the $\Delta E$ values of which deviate from experiment by 
2.18 MeV (1.66 MeV) and 2.16 MeV (1.2 MeV) in the mean field (PVC) calculations, 
respectively. We did not find clear explanation for such large differences.

  Proton and neutron pseudospin doublets with the same single-particle 
structure  have similar splitting energies in experiment (see pseudospin 
doublets indicated by the brackets with arrows in Fig.\ \ref{pseudo}). This 
feature is rather well reproduced both in the mean field and
PVC calculations for the proton and neutron $2p_{3/2}1f_{5/2}$ pseudospin
doublets of $^{56}$Ni. On the contrary, the mean field (``def+TO'') 
calculations completely fail to reproduce this feature for the proton
and neutron $2d_{5/2} 1g_{7/2}$ pseudospin doublets in $^{132}$Sn and 
$2f_{7/2} 1h_{9/2}$ pseudospin doublets in $^{208}$Pb (Fig.\ \ref{pseudo}); 
the calculated splitting energies for neutron doublets are by more than
1 MeV larger than the ones for proton doublets. However, in the PVC 
calculations the splitting energies for these pairs of proton and neutron
pseudospin doublets are similar in agreement with experiment. This is a 
consequence of the fact that particle-vibration coupling decreases 
(as compared with mean field calculations) the splitting energies in neutron 
pseudospin doublets by more than 1 MeV leaving at the same time the splitting 
energies in proton pseudospin doublets unchanged.

%%%%%%%%%%%%%%%%%%%%%%%%%%%%%%%%%%%%%%%%%%%%%%%%%%%%%%%%%%%%%%
\subsection{The $^{292}_{172}120$ nucleus - the center of the
island of stability of superheavy nuclei}
%%%%%%%%%%%%%%%%%%%%%%%%%%%%%%%%%%%%%%%%%%%%%%%%%%%%%%%%%%%%%%

   The superheavy $^{292}$120 nucleus has been predicted to 
have proton $Z=120$ and neutron $N=172$ spherical shell closures 
within the RMF theory \cite{BRRMG.99}. Thus, it represents
the center  of the "island of stability" of shell-stabilized 
superheavy nuclei. The analysis of the deformed one-quasiparticle
states in the $A\sim 250$ mass regions in Ref.\ \cite{AKFLA.03} 
supports the presence of the $Z=120$ shell gap in superheavy 
nuclei. In addition, it indicates $N=172$ as a likely candidate 
for magic neutron number in superheavy nuclei.

\begin{figure*}
\includegraphics[width=14.0cm]{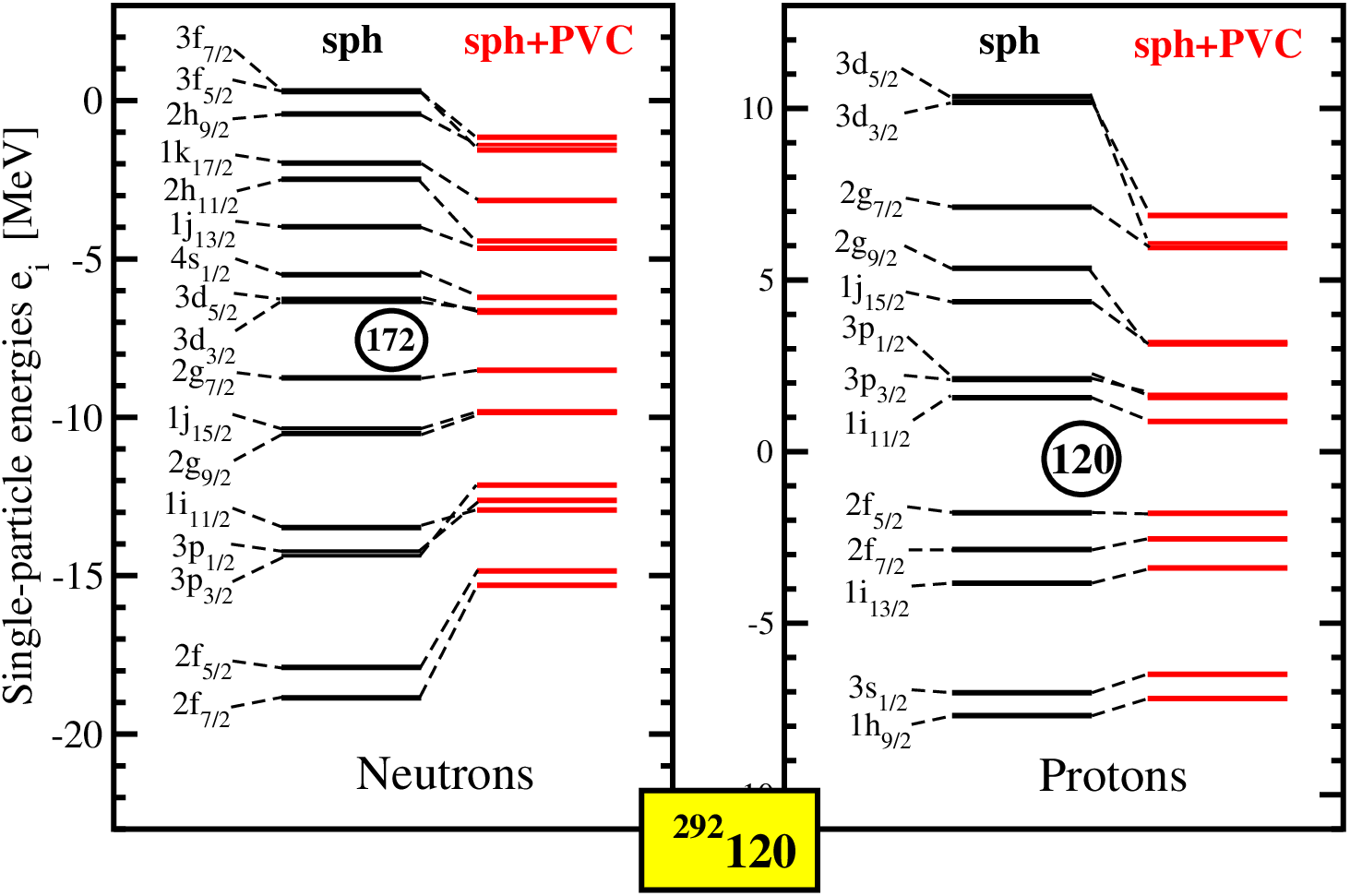}
\caption{(Color online) Single-particle spectra of the $^{292}120$ nucleus.
Column 'sph' shows the single-particle spectra obtained in spherical RMF 
calculations. Column 'sph+PVC' shows the spectra obtained in spherical 
calculations within the PVC model.}
\label{Fig-120-172}
\end{figure*}

   This nucleus represents a challenge for future experimental 
synthesis since it is located at the limits of accessibility 
by available cold fusion reactions.  Therefore, as accurate 
as possible estimations of its characteristics are needed 
from the theoretical side. First results for its single-particle 
spectra obtained within the relativistic particle-vibration 
coupling model are presented in Fig. \ref{Fig-120-172}. They 
are compared with the results obtained on the mean field level.
Note that we restrict ourselves in the case of the $^{292}$120 
nucleus to spherical calculations. This is because the polarization 
effects due to deformation and TO mean fields decrease with 
mass (see Sect.\ \ref{Pol-eff}), and thus their impact on the 
single-particle spectra is expected to be rather small in the 
$^{292}$120 nucleus.

  The density of the single-particle states is substantially larger
than in lighter nuclei at the mean field level. As a consequence,
this nucleus exhibits a very rich spectrum of low-lying collective
phonons already within the RRPA. For example, the lowest vibrational
$2^{+}_1$ mode appears at 1.41 MeV with B(E2)$\uparrow$ = 7.1
$\times 10^3$ e$^2$ fm$^4$, accompanied by several rather collective
modes at 3.18 MeV, 5.25 MeV, 6.74 MeV etc.
Similar picture is obtained for other multipolarities. Although
contributions to the nucleonic self-energy from the high-lying modes
decrease quickly with energy, we have included  phonons with all
J$^{\pi}$ values mentioned in Sec.\ \ref{Cut-off} below 15 MeV, altogether 
about 100 phonons.

  Fig.\ \ref{Fig-120-172} shows that the particle-vibration coupling 
gives rise to the general compression of the spectra. As a consequence,
the size of the $Z=120$ shell gap decreases from 3.35 MeV (in mean field
calculations) down to 2.67 MeV (in the PVC calculations), and the size 
of the neutron shell gap at $N=172$ from 2.42 MeV down to 1.83 MeV. 
Although these gaps are smaller by 20-25\% than the ones obtained in the 
mean field calculations, they are still significant. As a result, the 
$^{292}$120 nucleus still remains doubly magic nucleus representing the 
center of the ``island of stability'' of superheavy nuclei even when the
correlations beyond mean field are taken into account.

%%%%%%%%%%%%%%%%%%%%%%%%%%%%%%%%%%%%%%%%%%%%%%%%%%%%%%%%%%%%%%%%%%%%%%%%%%%%%%
%%%%%%%%%%%%%%%%%%%%%%%%%%%%%%%%%%%%%%%%%%%%%%%%%%%%%%%%%%%%%%%%%%%%%%%%%%%%%%

%%%%%%%%%%%%%%%%%%%%%%%%%%%%%%%%%%%%%%%%%%%%%%%
\section{Conclusions}
\label{Sec-final}
%%%%%%%%%%%%%%%%%%%%%%%%%%%%%%%%%%%%%%%%%%%%%%%

 The relativistic particle-vibration model has systematically been 
applied in combination with cranked relativistic mean field approach 
for the study of the impact of surface vibrations on single-particle 
motion. The polarization effects in odd mass nuclei due to deformation 
and time-odd mean fields have been treated in the CRMF  framework. 
The main results can be summarized as follows:

\begin{itemize}

\item
  Particle-vibration coupling has to be taken into account 
when model calculations are compared with experiment since this 
coupling is responsible for observed fragmentation 
of experimental levels. The inclusion of particle-vibration coupling
substantially improves the description of the energies of dominant 
single-particle states in $^{132}$Sn and $^{208}$Pb. However, the 
accuracy of the description of single-particle spectra in $^{56}$Ni 
is similar in the mean field and PVC calculations. Note that dominant 
neutron single-particle states are in average better described than 
proton  ones in the PVC calculations.

\item
The polarization effects in odd-mass nuclei due to deformation
and time-odd mean fields induced by odd particle are important. 
They have to be taken into account when experimental and calculated 
single-particle energies are compared. However, they are usually 
neglected when the results of particle-vibration coupling model 
calculations are compared with experiment. The current analysis 
indicates that within the framework of CDFT this neglect is more
or less justified only for heavy nuclei, and it is more justified 
for proton subsystem than for neutron one.

\item
 Particle-vibration coupling leads to a shrinkage of the shell 
gaps. The size of the shell gaps is typically underestimated in 
the PVC calculations as compared with experiment and overestimated 
in the mean field calculations.

\item
The inclusion of particle-vibration coupling decreases the accuracy 
of the description of spin-orbit splittings. The analysis suggests 
that the mean field constraint on the allowed range of the Dirac 
effective mass $0.57 \leq m^*_D/m \leq 0.61$ is too restrictive and 
the range of the $m^*_D/m$ values can be broader in the models which 
take into account the correlations beyond the mean field.

\item
  Particle-vibration coupling substantially improves the description 
of splitting energies in pseudospin doublets as compared with mean
field calculations. Observed similarity of the splitting energies of
proton and neutron pseudospin doublets with the same single-particle 
structure in medium and heavy mass nuclei can only be reproduced when 
the particle-vibration coupling is taken into account.

\item
  The spherical shell closures in superheavy nuclei are still 
found at proton $Z=120$ and neutron $N=172$ numbers even when 
particle-vibration coupling is taken into account. However, the 
size of these gaps becomes smaller (as compared with mean field 
values) in the presence of particle-vibration coupling due to 
general compression of the spectra caused by the increase of 
the effective mass of the nucleon at the Fermi level.

\item
  The NL3* parametrization employed in the current work has been
adjusted at the mean-field level. The remaining discrepancies 
between the results of particle vibration coupling model and 
experiment clearly  suggest that this parametrization is not 
completely adequate for the description of the energies of
the single-particle states.  We believe 
that this statement is not limited to NL3* but it is also valid 
for all existing CDFT parametrizations which have been adjusted 
at the mean-field level. This calls for the parametrizations 
specifically tailored to describe single-particle degrees of 
freedom in the models taking into account the correlations beyond the
mean field. 

\end{itemize}

%%%%%%%%%%%%%%%%%%%%%%%%%%%%%%%%%%%%%%%%%%%%%%%
\section{Acknowledgements}
%%%%%%%%%%%%%%%%%%%%%%%%%%%%%%%%%%%%%%%%%%%%%%%

 This work was supported by the LOEWE program of the State 
of Hesse (Helmholtz International Center for FAIR) and by 
the U.S. Department of Energy under the grant 
DE-FG02-07ER41459.

%%%%%%%%%%%%%%%%%%%%%%%%%%%%%%%%%%%%%%%%%%%%%%%%%%%%%%%%%%%%%%%%%%

\end{document}